\documentclass[12pt]{iopart}
\usepackage{color}
\usepackage{cite}
\usepackage{braket}
\usepackage[pdftex]{graphicx}
\usepackage{iopams}  

\makeatletter
\def\@makefnmark{\hbox{\textsuperscript{\normalfont\@thefnmark}}}
\let\@fnsymbol\@arabic
\makeatother

\begin{document}

\title[]{Improved systematic uncertainty evaluation of the $^{171}$Yb optical lattice clock NMIJ-Yb1 with uncertainty of $2.6\times10^{-17}$}

\author{Takumi Kobayashi$^{1,*}$, Akiko Nishiyama$^{1}$, Ikuhiko Saito$^{1}$, Daisuke Akamatsu$^{2}$, Akio Kawasaki$^{1}$, Shintaro Nagase$^{1}$, Masami Yasuda$^{1}$}

\address{$^1$National Metrology Institute of Japan (NMIJ), National Institute of Advanced Industrial Science and Technology (AIST), 1-1-1 Umezono, Tsukuba, Ibaraki 305-8563, Japan\\
$^2$Department of Physics, Graduate School of Engineering Science, Yokohama National University, 79-5 Tokiwadai, Hodogaya-ku, Yokohama 240-8501, Japan\\}
\ead{takumi-kobayashi@aist.go.jp}
\vspace{10pt}
%\begin{indented}
%\item[\today]
%\end{indented}

\begin{abstract}
We report a systematic uncertainty evaluation of the $^{171}$Yb optical lattice clock NMIJ-Yb1 (NMIJ: National Metrology Institute of Japan) with a fractional frequency uncertainty of $2.6\times10^{-17}$, improved by a factor of 3.8 compared with our previous uncertainty. The uncertainty of the lattice light shift is reduced to $1.6\times10^{-17}$ by preparing axially sideband-cooled atoms in a shallow optical lattice. Different models to calculate the lattice light shift are compared, taking account of possible model-dependent biases due to the treatment of the radial motion of atoms trapped in the optical lattice. The uncertainty of the blackbody radiation shift reaches $7.0\times10^{-18}$ by measuring the radiative temperature at the position of atoms with an in-vacuum temperature sensor. We also demonstrate the nearly continuous operation of NMIJ-Yb1 with an uptime of 91.3 $\%$ for 10 days, showing the potential for future improvement of the calibration uncertainty of International Atomic Time.
\end{abstract}

%\maketitle

\section{Introduction}
There has been an increasing number of reports on optical clocks such as optical lattice clocks and single ion optical clocks with fractional frequency uncertainties at the $10^{-18}$ to $10^{-19}$ level \cite{Ushijima2015,Hunteman2016,McGrew2018,Brewer2019,Bothwell2019,Tofful2024,Li2024,Aeppli2024,Lu2025,Nosske2025,Lindvall2025pra,Jia2026,Zhang2026,ZhangB2026,Zhu2026}. With the remarkable advances, optical clocks are considered as sensitive quantum sensors for measuring the geopotential \cite{Takano2016,Grotti2018,Grotti2024,Liu2024} and searching for physics beyond the standard model \cite{Wcislo2018,Takamoto2020,Kennedy2020,Lange2021,Kobayashi2022,Filzinger2023,Kawasaki2025} as well as candidates for a redefinition of the second in the International System of Units (SI) \cite{Hong2016,Dimarcq2024}. 

Towards the redefinition of the SI second, the Consultative Committee for Time and Frequency (CCTF) has determined mandatory criteria that must be achieved before the redefinition \cite{Dimarcq2024}. Some of the criteria have already been achieved with high fulfilment levels (e.g., continuity with the definition based on Cs) thanks to great effort made by many laboratories worldwide. Nevertheless, critical issues are currently recognized for two criteria: the validation of uncertainty budgets of optical clocks and the contribution of optical clocks to International Atomic Time (TAI) \cite{CGPM2026}. Regarding the validation of uncertainty budgets, agreement of the frequency ratios of optical clocks measured several times by different institutes with uncertainties of $\lesssim5\times10^{-18}$ is required. While some groups have reported the frequency ratios at the low $10^{-18}$ level \cite{McGrew2018,Takamoto2020,Sanner2019,Aeppli2025,Hausser2025}, agreement between measurements by different institutes has not been reported at the required level. In fact, recent international clock comparisons have revealed disagreement of the frequency ratios even at the $10^{-17}$ to $10^{-16}$ level \cite{Lindvall2025,Aeppli2025,Amy-Klein2024,Dawel2026,Pizzocaro2026}, indicating the importance of reevaluation of the uncertainty budget and participation in clock comparisons from many groups. Regarding the contribution to TAI, at least three frequency calibrations of TAI with uncertainties of $\lesssim2\times10^{-16}$ each month from a set of at least five optical clocks for at least 1 year are required. So far, only limited TAI calibrations have achieved this uncertainty level, and the number of calibrations has not reached the required level \cite{circulart}.

The $^{171}$Yb optical lattice clock NMIJ-Yb1 (NMIJ: National Metrology Institute of Japan) has contributed to TAI as a secondary frequency standard (SFS) that is officially allowed to participate in the TAI calibration. The unique feature of NMIJ-Yb1 is the capability of the operation with a high uptime (e.g, $> 80$ $\%$ uptime for 6 months \cite{Kobayashi2020}), which enables us to reduce the uncertainty of the link between NMIJ-Yb1 and TAI to $\lesssim2\times10^{-16}$ for 30 days. However, the systematic uncertainty of NMIJ-Yb1 is so far $\sim1\times10^{-16}$ \cite{Kobayashi2022}, which is not small enough to achieve the criterion for the TAI calibration. 

In this work, we report an improved systematic uncertainty evaluation of NMIJ-Yb1 with an uncertainty of $2.6\times10^{-17}$ mainly for achieving the criterion for the contribution to TAI. Major improvements are made in (i) the lattice light shift by preparing sideband-cooled atoms in a shallow optical lattice and (ii) the blackbody radiation (BBR) shift by measuring the radiative temperature at the position of atoms with an in-vacuum temperature sensor. We also demonstrate the operation of NMIJ-Yb1 with $>90$ $\%$ uptime for 10 days, showing the potential for future improvement of the TAI calibration. Furthermore, the present uncertainty at the low $10^{-17}$ level can contribute to future clock comparisons for finding the reason for the internationally observed disagreement of the frequency ratios at the $10^{-17}$ to $10^{-16}$ level, which is an essential step towards the achievement of the criterion for validating the uncertainty budget. 

\section{Experimental setup}
\label{experimentalsetupsection}
\begin{figure}[h]
\includegraphics[scale=0.6]{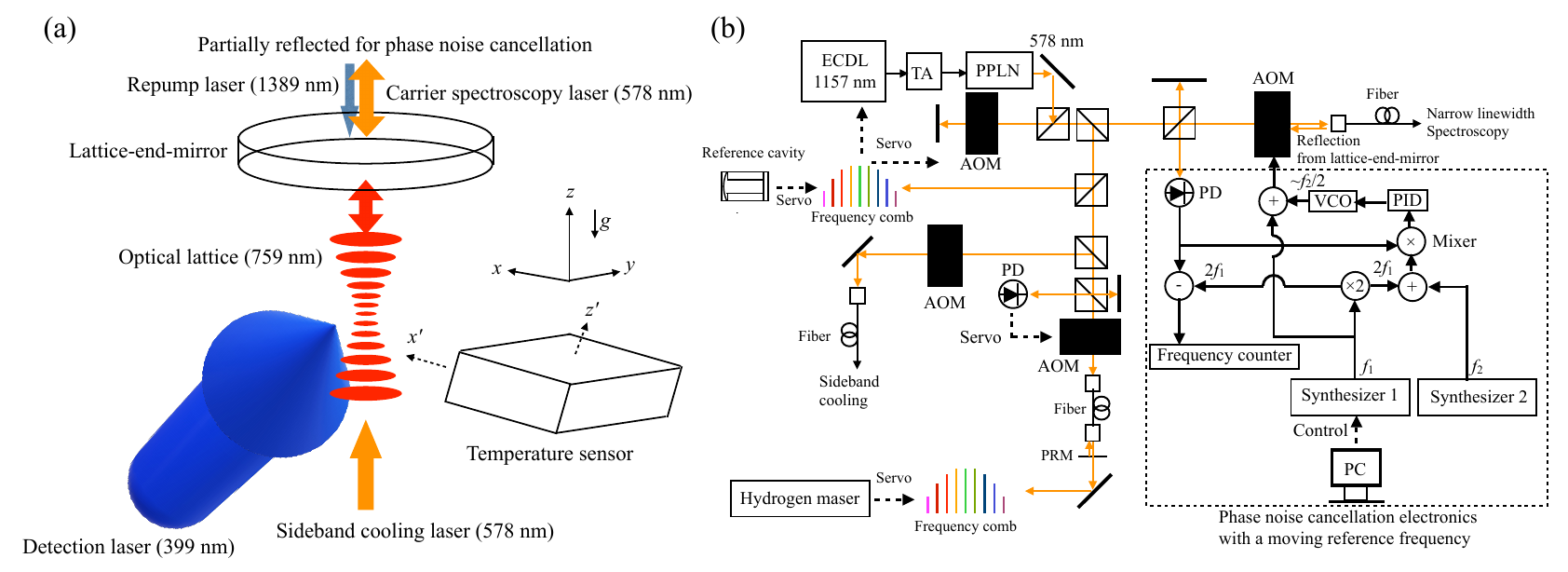}
\caption{(a) Schematic diagram of the experimental setup highlighting key updates (not to scale). (b) 578 nm laser system. ECDL: External cavity diode laser, TA: Tapered amplifier, PPLN: periodically poled LiNbO$_{3}$, AOM: Acousto-optic modulator, PD: Photodetector, PRM: Partially reflecting mirror, VCO: Voltage controlled oscillator, PID: Proportional integral derivative, PC: Personal computer.}
\label{experimentalsetup}
\end{figure}

Figure \ref{experimentalsetup} (a) shows a schematic diagram of the experimental setup. We here describe key updates, wheares the details of our setup are given elsewhere \cite{Kobayashi2018,Kobayashi2020}. After the first- and second-stage cooling on the $^{1}$S$_{0}$-$^{1}$P$_{1}$ and $^{1}$S$_{0}$-$^{3}$P$_{1}$ transitions, respectively, Yb atoms are loaded into a vertically oriented 1-D optical lattice operated at the magic wavelength of 759 nm. We label the vertical axis $z$ and the radial axes $x$ and $y$. To evaluate the BBR shift, we insert an in-vacuum temperature sensor which moves along $x^{\prime}$ axis tilted by $18^{\circ}$ about the $y$ axis from the $x$ axis. For uncertainty evaluation described in Section \ref{systematicuncertaintysection}, we also define $z^{\prime}$ axis tilted by $18^{\circ}$ about the $y$ axis from the $z$ axis. For detecting trapped atoms, a 399-nm laser resonant to the $^{1}$S$_{0}$-$^{1}$P$_{1}$ transition is irradiated in the direction tilted by $135^{\circ}$ about the $y$ axis from the $x$ axis.

We implement axial sideband cooling \cite{Nemitz2016} along the $z$ direction by the 578-nm and 1389-nm repump lasers resonant to the $^{1}$S$_{0}$-$^{3}$P$_{0}$ and $^{3}$P$_{0}$-$^{3}$D$_{1}$ transitions, respectively. The 578-nm laser induces the red sideband transition, driving atoms from higher to lower motional states. Atoms in the excited $^{3}$P$_{0}$ state are then back to the ground $^{1}$S$_{0}$ state via the $^{3}$P$_{0}$-$^{3}$D$_{1}$ transition induced by the 1389-nm laser. This cycle is repeated by simultaneously irradiating the 578-nm and 1389-nm lasers for 60 ms, accumulating atoms in the motional ground state. The sideband cooling is carried out 50 ms after atoms are loaded to the lattice at a full lattice laser power, followed by spin polarization by the $^{1}$S$_{0}$-$^{3}$P$_{1}$ transition and reduction of the lattice laser power to a target value (see Section \ref{latticelightsection} for typical trap depths of the optical lattice). Afterwards, narrow linewidth spectroscopy of the carrier transition is performed by irradiating another 578-nm laser for 40 ms along the $z$ direction from the opposite direction of the sideband cooling laser. To suppress the first-order Doppler shift, the 578-nm laser for the carrier spectroscopy is partially reflected by a lattice-end-mirror for stabilizing the optical path length \cite{Falke2012}. 

Figure \ref{experimentalsetup} (b) shows the 578-nm laser system. To obtain sufficient power for the sideband cooling, carrier spectroscopy, and comb-based frequency comparison with a hydrogen maser, we insert a tapered amplifier for an external cavity diode laser (ECDL) at the fundamental wavelength of 1157 nm. The ECDL is phase-locked to an optical frequency comb \cite{Inaba2013} which is stabilized against an ultra-stable reference optical cavity. In the evaluation of the lattice light shift and density shift (see Section \ref{systematicuncertaintysection}), we employ a recently installed ultralow-expansion (ULE) cavity with crystalline mirrors at 1560 nm \cite{Nishiyama} as a reference cavity, while our conventional ULE cavity at 1064 nm \cite{Kobayashi2018} is utilized for the TAI calibration (see Section \ref{discussionsection}). All three functions of frequency steering, power stabilization, and optical path length stabilization for the carrier spectroscopy laser are performed using a single acousto-optic modulator (AOM). For the optical path length stabilization, a Michelson interferometer detects the phase noise experienced by the 578-nm light on the path to the lattice-end-mirror via an optical fiber. The phase noise cancellation is applied by phase-locking the beat note detected in the interferometer to a reference synthesizer, and is checked by counting the beat note with a frequency counter. Since the frequency of the beat note is changed by steering the AOM frequency to lock the laser to the $^{1}$S$_{0}$-$^{3}$P$_{0}$ transition, it is impossible in a conventional method to count a constant frequency to monitor cycle slipping events. To achieve this, we (i) generate the reference frequency for the phase locking by adding two RF frequencies ($2f_{1}+f_{2}$ in Fig.~\ref{experimentalsetup} (b)), one of which is a frequency-doubled output of a synthesizer (Synthesizer 1) used for steering the AOM at a frequency of $f_{1}+f_{2}/2$, and (ii) subtract $2f_{1}$ from the beat note frequency. The resulting constant frequency is fed to the frequency counter\footnote[1]{We have already employed this scheme in previous measurements \cite{Kobayashi2022,Kobayashi2025}, while not explicitly described.}. Regarding the stabilization of the optical path length to another frequency comb \cite{Inaba2006} which is phase-locked to the hydrogen maser, we implement a conventional phase noise cancellation scheme with an AOM at a constant frequency. 

\section{Systematic uncertainty evaluation}
\label{systematicuncertaintysection}
Table \ref{systematictalbe} lists the systematic frequency shifts and uncertainties of NMIJ-Yb1. The total uncertainty is $2.6\times10^{-17}$ without the uncertainty of the gravitational redshift \cite{Nakashima2022}.
 
\begin{table}[h]
\caption{Updated systematic uncertainty budget of NMIJ-Yb1. The frequency shifts and uncertainties are given by fractional values relative to the $^{1}$S$_{0}$-$^{3}$P$_{0}$ transition frequency of $^{171}$Yb (518 THz). This budget is evaluated for the 10-day measurement described in Section \ref{nearlycontinuoussection}. Note that the budget published for the TAI evaluation \cite{circulart} has not been updated.}  
%	\legend{\small{Only positive total $M$ values are shown. The table is symmetric for $M<0$.}}
	\label{systematictalbe}
	\begin{center} 
\begin{tabular}{lcc}
\hline
Effect  & $\mathrm{Shift}$ $(\times10^{-18}$ Hz/Hz) & $\mathrm{Uncertainty}$ $(\times10^{-18}$ Hz/Hz) \\
\hline
Lattice light & 44.2 & 15.8 \\
BBR & $-2545.2$ & 7.0 \\
Density & $-3.4$ & 5.5 \\
Second-order Zeeman & $-54.9$ & 3.0 \\
Probe light & 4.0 & 9.5 \\
Servo error & $54.9$ & 15.1\\
AOM switching & $-$& 0.1 \\
Line pulling & $-$& 2.0 \\
DC Stark & $-$& $3.0$ \\
Background gas & $-2.7$ & 0.8 \\
\hline
Total & $-2503.2$ & 25.9$$\\
\hline
Gravitational redshift &2307.8 &5.6\\
\hline
Total (with gravitational redshift) &$-195.4$ &26.5\\
\hline
\end{tabular}
\end{center}
\end{table}

\subsection{Lattice light shift}
\label{latticelightsection}
We follow a light shift model based on the Born-Oppenheimer + WKB (semiclassical) approximation \cite{Beloy2020}. Without radial cooling \cite{Zhang2022,Chen2024} in our experimental setup, we expect that this model yields a better treatment of the radial motion of atoms compared with other models utilizing the harmonic oscillator approximation \cite{Katori2015,Ushijima2018,Brown2017}. For comparison, we also carry out another analysis with the harmonic oscillator model as described in Section \ref{discussionsection}. The Born-Oppenheimer + WKB model gives the lattice light shift $\Delta \nu_{\mathrm{LS}}$ due to the electric-dipole ($E1$) polarizability, multipolar ($M1$ and $E2$) polarizabilities, and hyperpolarizability by 
\begin{eqnarray}
\Delta \nu_{\mathrm{LS}}&=&-\sum_{n_{z}}P_{n_{z}}\Big[\alpha_{E1}\Delta\nu X(n_{z},U_{\mathrm{max}},T_{\mathrm{r}})(U_{\mathrm{max}}/E_{\mathrm{r}})\nonumber\\
&&+\alpha_{M1E2}Y(n_{z},U_{\mathrm{max}},T_{\mathrm{r}})(U_{\mathrm{max}}/E_{\mathrm{r}})\nonumber\\
&&+\beta Z(n_{z},U_{\mathrm{max}},T_{\mathrm{r}})(U_{\mathrm{max}}/E_{\mathrm{r}})^{2}\Big],
\label{bolightshifteq}
\end{eqnarray}
where $P_{n_{z}}$ is the fractional population at the state with the axial quantum number $n_{z}$, $\Delta \nu$ the detuning of the lattice laser frequency $\nu_{\mathrm{L}}$ from the $E1$ magic frequency $\nu_{E1}$, $U_{\mathrm{max}}$ the maximum trap depth, and $T_{\mathrm{r}}$ the radial temperature. $E_{\mathrm{r}}=h^{2}\nu_{\mathrm{L}}^{2}/2mc^{2}$ is the lattice photon recoil energy, $h$ the Planck constant, $c$ the speed of light, and $m$ the mass of the atom. $X(n_{z},U_{\mathrm{max}},T_{\mathrm{r}})$, $Y(n_{z},U_{\mathrm{max}},T_{\mathrm{r}})$, and $Z(n_{z},U_{\mathrm{max}},T_{\mathrm{r}})$ are reduction factors of the trap depth, which takes account of the radial motion of atoms. A detailed description of the reduction factors is given in Ref.~\cite{Beloy2020}, but we also put their forms in Appendix A. We calculate the reduction factors by numerical integration with a code written in Mathematica. $\alpha_{E1}$, $\alpha_{M1E2}$, and $\beta$ are the $E1$ polarizability, multipolar polarizability, and hyperpolarizability coefficients, respectively. We take weighted mean values $\alpha_{E1}=23.59(1.41)$ $\mathrm{\mu}$Hz/MHz, $\alpha_{M1E2}=-736(47)$ $\mathrm{\mu}$Hz, and $\beta=-1098(60)$ $\mathrm{\mu}$Hz of the experimentally determined coefficients \cite{Nemitz2019,Kim2021,Bothwell2025,Pizzocaro2020,Kobayashi2018}. Regarding $\alpha_{E1}$, the uncertainty is inflated by the square root of the reduced chi-squared $\sqrt{\chi_{\mathrm{red}}^{2}}$. 

\begin{figure}[h]
\includegraphics[scale=0.45]{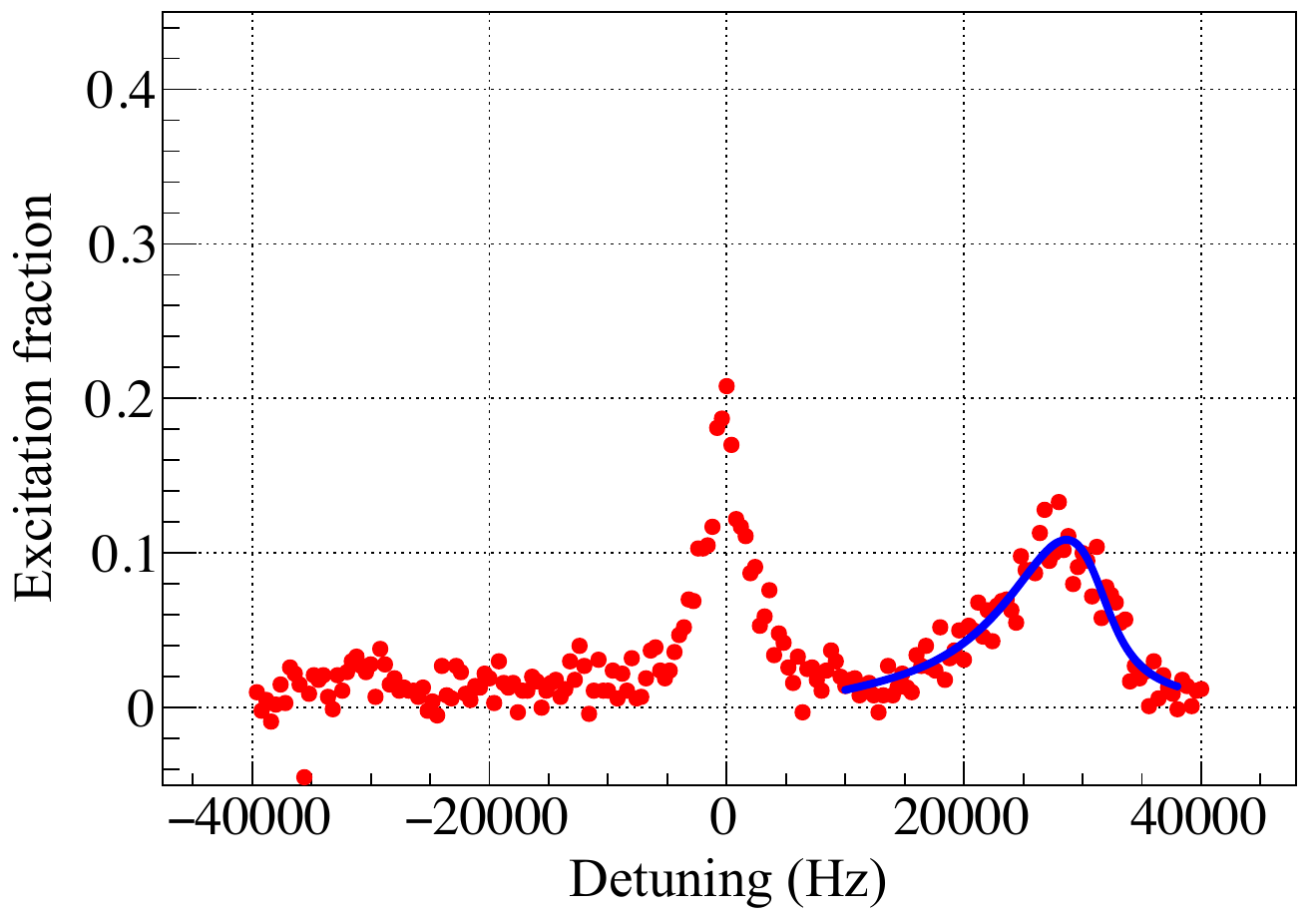}
\caption{Sideband spectrum of the $^{1}$S$_{0}$-$^{3}$P$_{0}$ transition obtained with axial sideband cooling. The blue line indicates a fit function given by Eq.~(\ref{sidebandfitfunc}) to extract $U_{\mathrm{max}}$ and $T_{\mathrm{r}}$ based on the Born-Oppenheimer approximation model \cite{Beloy2020}.}
\label{sidebandfigure}
\end{figure}

To determine the parameters $P_{n_{z}}$, $U_{\mathrm{max}}$, and $T_{\mathrm{r}}$, we perform sideband spectroscopy of the clock transition. Figure \ref{sidebandfigure} shows a sideband spectrum obtained in a typical operational condition. From the ratio of the red and blue sidebands, $P_{0}$ is estimated as $\gtrsim80$ $\%$ achieved by the sideband cooling. Assuming that the remaining population is predominantly in the $n_{z}=1$ state ($P_{1}\lesssim20$ $\%$), we restrict ourselves to the $n_{z}\le1$ states in the present analysis. For the estimation of $U_{\mathrm{max}}$ and $T_{\mathrm{r}}$, we employ a model of the trap potential $U_{n_{z}}(\rho)$ for $n_{z}$ as a function of the radial coordinate $\rho$ based on the Born-Oppenheimer approximation \cite{Beloy2020}. The form of $U_{n_{z}}(\rho)$ is found in Appendix A. Following previous analyses \cite{Bothwell2025,Goti2025}, we include $U_{n_{z}}(\rho)$ in a fit function for the blue sideband spectrum. First, an evenly sampled set of $\sim100$ frequencies $\nu_{\mathrm{SB}}^{n_{z},k}$ ($k$ is from 1 to $\sim100$) are assigned for the blue sideband transition $n_{z}\to n_{z}+1$ from $\sim10$ kHz to the corner frequency at the high frequency edge. Each frequency $\nu^{n_{z},k}_{\mathrm{SB}}$ is then related to the radial position $\rho_{n_{z},k}$ by $h\nu^{n_{z},k}_{\mathrm{SB}}=U_{n_{z}+1}(\rho_{n_{z},k})-U_{n_{z}}(\rho_{n_{z},k})$. The lineshape of each sideband transition with $\nu_{\mathrm{SB}}^{n_{z},k}$ is modeled by a Lorentzian function $\Omega^{2}(n_{z},\rho_{n_z,k})/[\Omega^{2}(n_{z},\rho_{n_z,k})+(\nu^{n_{z},k}_{\mathrm{SB}}-\nu_{\mathrm{cl}})^{2}]$ where $\nu_{\mathrm{cl}}$ is the 578-nm laser frequency and $\Omega(n_{z},\rho_{n_{z},k})$ the Rabi frequency for the blue sideband transition. The Rabi frequency is given by \cite{Leibfried2003}
\begin{equation}
\Omega(n_{z},\rho_{n_{z},k})=\Omega_{0}e^{-\eta^2(n_{z},\rho_{n_z,k})/2}\eta(n_{z},\rho_{n_z,k})\sqrt{\frac{n_{z}!}{(n_{z}+1)!}}L^{1}_{n_{z}}(\eta^2(n_{z},\rho_{n_z,k})),
\label{rabifrequencyeq}
\end{equation}
where  $\Omega_{0}$ is the bare Rabi frequency, $L_{n}^{\alpha}(x)$ the generalized Laguerre polynomial, and $\eta^{2}(\rho_{n_z,k})=E_{\mathrm{r}}^{\mathrm{cl}}/(2\sqrt{U_{0}(\rho_{n_z,k})E_{\mathrm{r}}})$ ($E_{\mathrm{r}}^{\mathrm{cl}}$: 578-nm photon recoil energy) the Lamb-Dicke parameter. Finally, the fit function $F$ for the sideband spectrum is obtained by a sum of the Lorentzian functions,
\begin{eqnarray}
F(\nu_{\mathrm{cl}},U_{\mathrm{max}},\Omega_{0},T_{\mathrm{r}},a,b)=&&\nonumber\\
\quad\quad\quad\quad a\sum_{n_{z}=0}^{1} P_{n_{z}} \sum_{k=1}^{\sim100}\frac{\rho_{n_z,k}^{2}e^{-(U_{n_z}(\rho_{n_z,k})-U_{n_z}(0))/k_{\mathrm{B}}T_{\mathrm{r}}}\Omega^{2}(n_z,\rho_{n_z,k})}{\Omega^{2}(n_z,\rho_{n_z,k})+(\nu^{n_z,k}_{\mathrm{SB}}-\nu_{\mathrm{cl}})^{2}}+b,
\label{sidebandfitfunc}
\end{eqnarray}
where $k_{\mathrm{B}}$ is the Boltzmann constant. In Eq.~(\ref{sidebandfitfunc}), we take account of the density of states \cite{Beloy2020} and Boltzmann weight for each $\rho_{n_z,k}$. In the fit of the sideband spectrum, the parameters $\Omega_{0}$, $T_{\mathrm{r}}$, $a$, and $b$ are treated as free parameters, while $U_{\mathrm{max}}$ is initially treated as a fixed parameter determined by the corner frequency. The fit is repeated by changing $U_{\mathrm{max}}$ around the initial value until the standard deviation of the fit residuals reaches a minimum value. The obtained fit curve is shown in Fig.~\ref{sidebandfigure} as a blue line. 

\begin{table}[t]
\caption{Trap parameters in the interleaved measurement between deep and shallow trap depths with a fixed lattice frequency.} 
%	\legend{\small{Only positive total $M$ values are shown. The table is symmetric for $M<0$.}}
	\label{botrapparametertable}
	\begin{center} 
\begin{tabular}{lccccccc}
\hline
 &$n_{z}$ & $U_{\mathrm{max}}/E_\mathrm{r}$ &    $P_{n_z}$ &$T_{\mathrm{r}}$ ($\mu$K) & $X(n_{z},U_{\mathrm{max}},T_{\mathrm{r}})$ & $Y(n_{z},U_{\mathrm{max}},T_{\mathrm{r}})$ & $Z(n_{z},U_{\mathrm{max}},T_{\mathrm{r}})$ \\
 \hline
Deep & 0 & 381(23) & 0.814(60) & 4.8(1.0)  & 0.815(39)& 0.0233(6) & 0.690(53)\\
&1 & 381(23) &  0.186(60) & 4.8(1.0)   & 0.758(37)& 0.0695(18)& 0.604(48)\\
\hline
Shallow & 0 & 76.0(4.3) & 0.902(39) &1.08(22) & 0.772(28)& 0.0522(7)& 0.621(42)\\
 &1 & 76.0(4.3) &  0.098(39) & 1.08(22) & 0.683(10)& 0.162(3)& 0.480(17)\\
 \hline
\end{tabular}
\end{center}
\end{table}

\begin{figure}[h]
\includegraphics[scale=0.5]{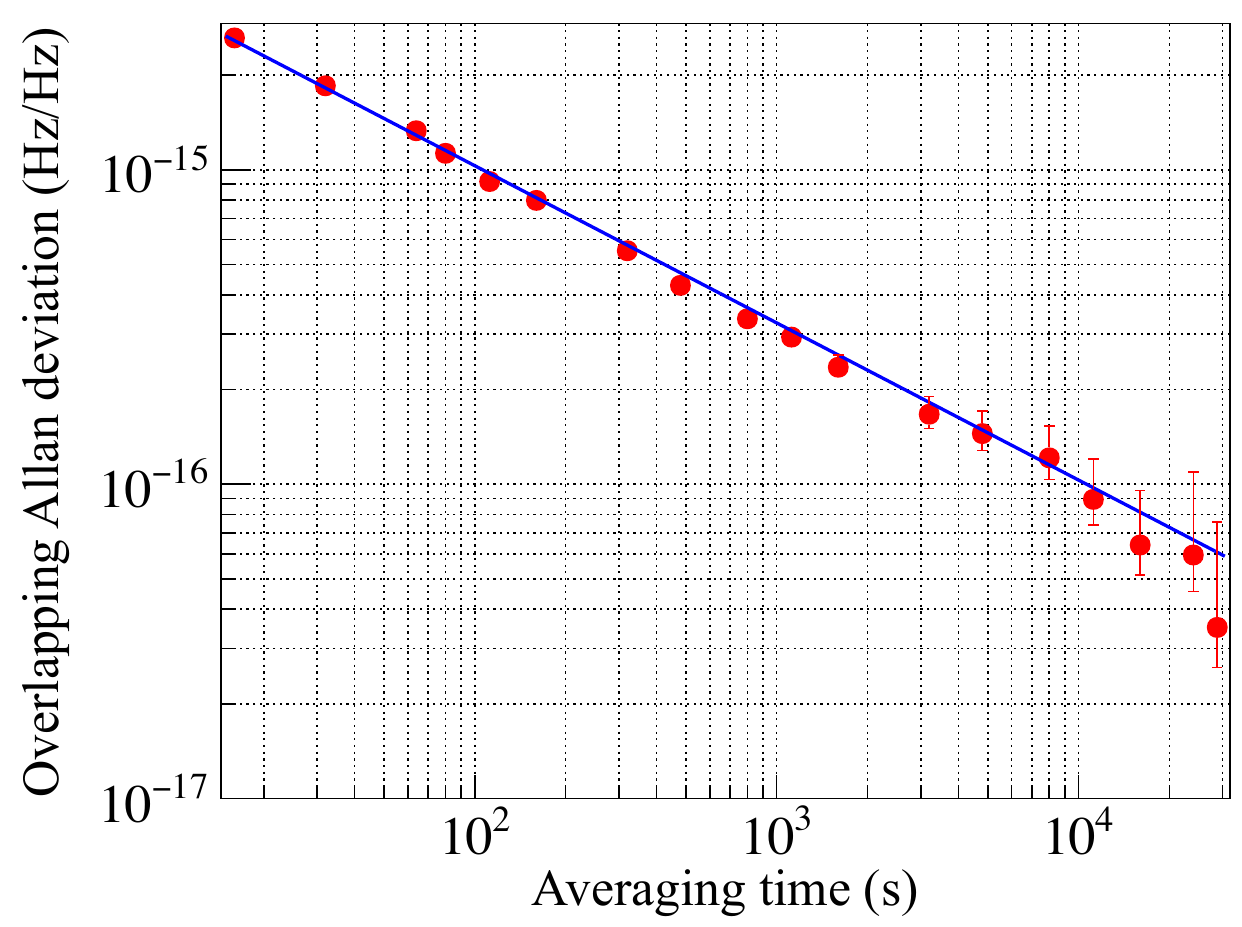}
\caption{Overlapping Allan deviation of the interleaved measurement as a function of the averaging time $\tau$. Solid blue line indicates a slope of $1.0\times10^{-14}/\sqrt{(\tau/\mathrm{s})}$.}
\label{interleavefig}
\end{figure}

To find $\nu_{E1}$ in our apparatus, we carry out an interleaved measurement in which the light shift at our operating lattice frequency $394\,798\,249.7$ MHz is measured by the alternative stabilization of the clock laser to the atomic transition with deep and shallow trap depths. The lattice is generated by a titanium sapphire laser, and its background spectrum from amplified spontaneous emission is filtered out by a volume Bragg grating with a bandwidth of 10 GHz \cite{Fasano2021}. The parameters in this measurement are summarized in Table \ref{botrapparametertable}. The statistical uncertainties of the parameters are determined by typical reproducibilities of the trap depth and temperature in our previous many measurements for the TAI calibration \cite{circulart}. We also include the systematic uncertainties arising from observed background excitations in the sideband spectrum at large detuning frequencies $\gtrsim100$ kHz, which is attributed to a limited servo bandwidth of the phase locking of the 578-nm laser \cite{Kobayashi2019}. The frequency stability of the interleaved measurement is shown in Figure \ref{interleavefig}. With an averaging time of $\tau\sim1$ day, we observe a frequency shift of $33.5(5.0)\times10^{-17}$, where the density shift is corrected according to the evaluation result in Section \ref{densityshiftresult}. The observed frequency shift yields $\nu_{E1}=394\,798\,261.0(5.8)$ MHz by Eq.~(\ref{bolightshifteq}) with uncertainty propagation by a Monte-Carlo method. The uncertainty of $\nu_{E1}$ is mostly due to the statistical uncertainty of the interleaved measurement. Our estimated $\nu_{E1}$ is in agreement with previous values reported by other groups \cite{Nemitz2019,Pizzocaro2020,Bothwell2025,Kim2021}. The uncertainty of the lattice light shift in an operating depth of $\sim70E_{\mathrm{r}}$ is estimated as $1.6\times10^{-17}$.

\subsection{Blackbody radiation shift}
The BBR shift $\Delta\nu_{\mathrm{BBR}}$ caused by an environment with a temperature $T_{\mathrm{atom}}$ is given by
\begin{equation}
\Delta \nu_{\mathrm{BBR}}=\alpha_{\mathrm{stat}}T^{4}_{\mathrm{atom}}+\alpha_{\mathrm{dyn,6}}T^{6}_{\mathrm{atom}}+\alpha_{\mathrm{dyn,8}}T^{8}_{\mathrm{atom}},
\label{bbrshifteq}
\end{equation}
where $\alpha_{\mathrm{stat}}$ is the coefficient arising from the static polarizability, and $\alpha_{\mathrm{dyn,6}}$ and $\alpha_{\mathrm{dyn,8}}$ are the dynamic corrections \cite{Beloy2014}. The values of the coefficients are taken from Refs.~\cite{Sherman2012,Hassan2025}. 

\begin{figure}[h]
\includegraphics[scale=0.45]{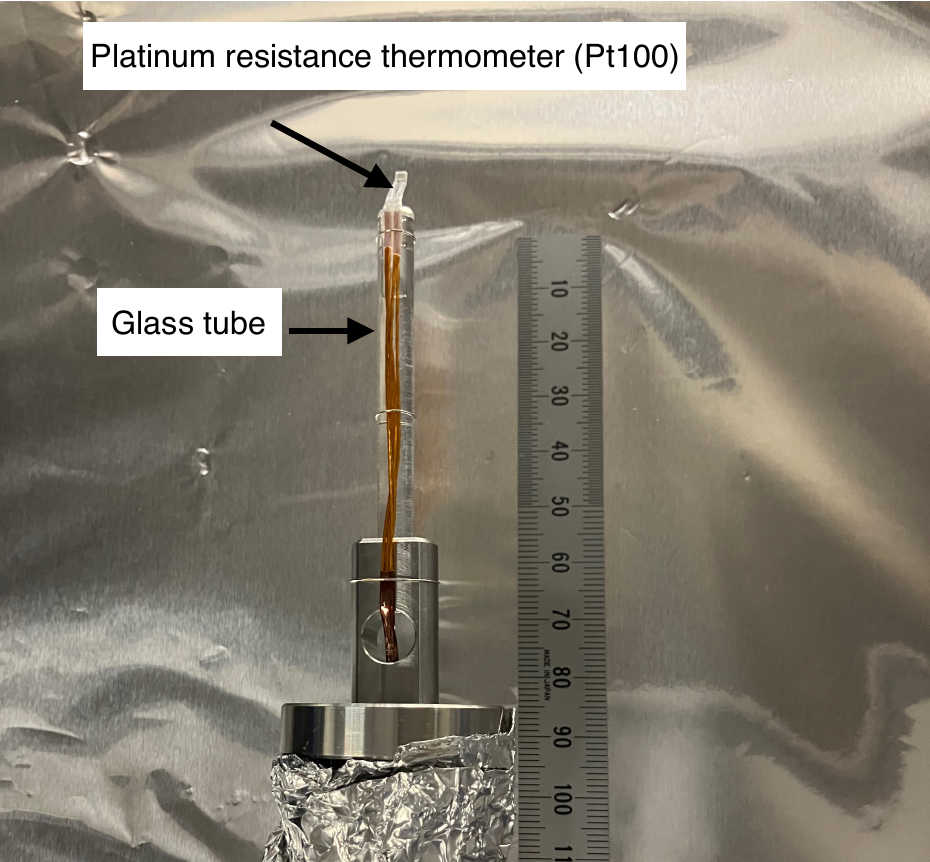}
\caption{Picture of the in-vacuum temperature sensor mounted on a glass tube.}
\label{sensorpicture}
\end{figure}

To precisely determine $T_{\mathrm{atom}}$, we directly measure the radiative temperature at the position of atoms with the in-vacuum movable temperature sensor (see Fig.~\ref{experimentalsetup}), inspired by previous studies \cite{Bothwell2019,Heo2022,Jin2023}. The sensor is a platinum resistance thermometer (Pt100) calibrated with an uncertainty of 3.0 mK. Its calibration is performed using the triple point of water (0.01 $^{\circ}$C) and the melting point of gallium (29.7646 $^{\circ}$C). Furthermore, the calibration uncertainty is evaluated from comparison measurements in a water bath with a standard platinum resistance thermometer traceable to the International Temperature Scale of 1990 \cite{ITS90} as realized at NMIJ. During the calibration and uncertainty evaluation, the resistance of the sensor is measured using a 4-wire configuration with an ISOTECH microK 250 precision thermometry bridge referenced to a 100 $\Omega$ standard resistor. We mount the sensor on a 60-cm-long glass tube, as shown in Figure \ref{sensorpicture}, to reduce the immersion error due to conductive heat transfer between the room temperature environment and the sensor. The sensor is positioned at the atom position with a linear motion micrometer, with the 399-nm detection and 578-nm sideband cooling laser beams serving as guides (see Fig.~\ref{experimentalsetup}). The radiative temperature is measured in the same condition as the clock operation. In the radiative temperature measurement, the resistance of the sensor is measured by a 4-wire multimeter (Lakeshore model 224) which is calibrated against a variable standard resistor (IET Labs RTD-Z-6-0.001) with a fractional uncertainty of $5\times10^{-7}$ $\Omega/\Omega$, corresponding to a temperature uncertainty of 15 mK. 

\begin{figure}[h]
\includegraphics[scale=0.55]{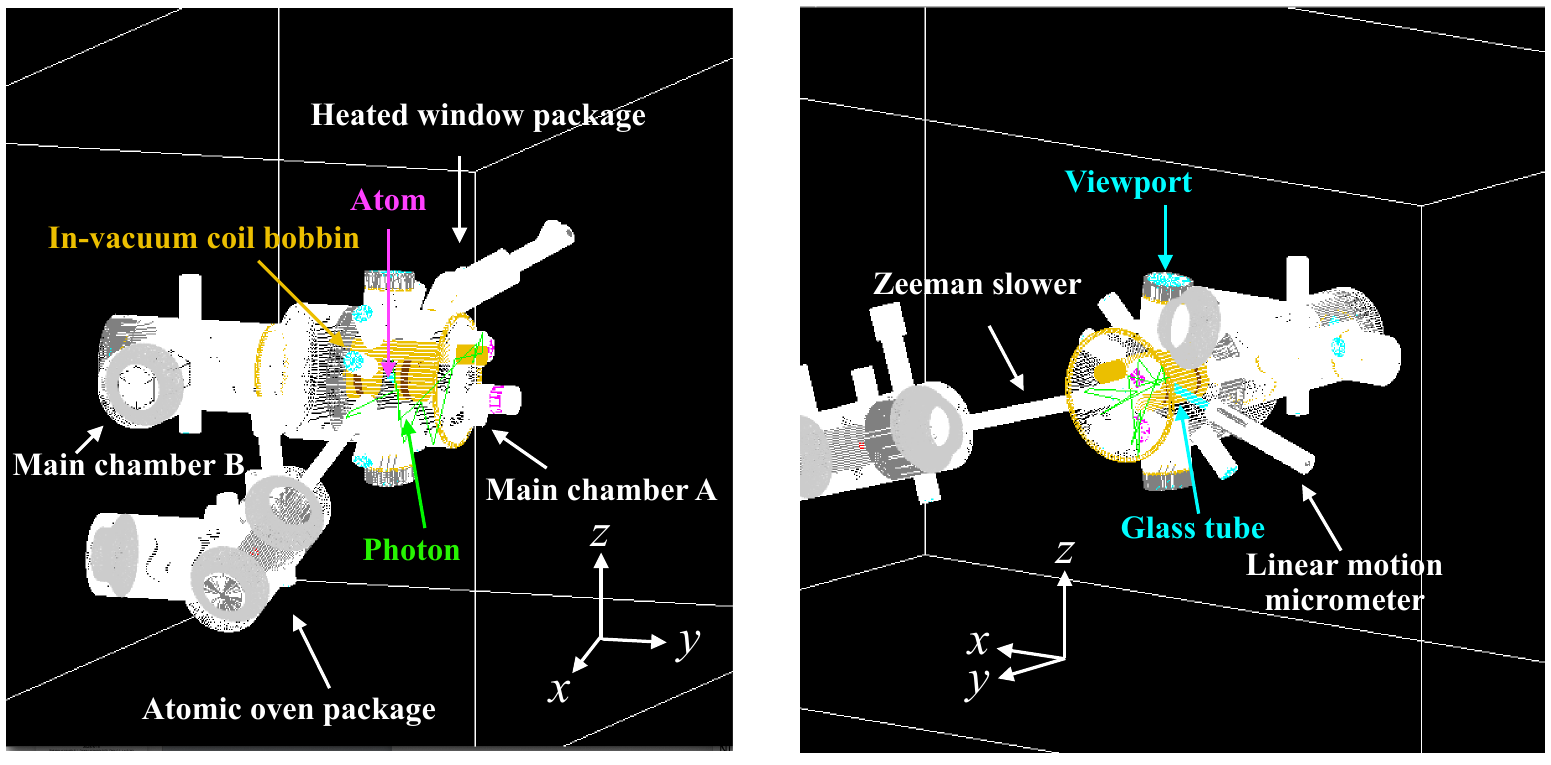}
\caption{Chamber geometries seen from different angles for the Monte-Carlo simulation of the BBR photon trajectory (green line) based on the Geant4 simulation toolkit \cite{Allison2016}. The initial design of the chamber is made by RIKEN \cite{Takamoto2020,Ohmae2021}.}
\label{geant4geometry}
\end{figure}

Since the in-vacuum sensor is retracted away from the atom position during the clock operation, we construct a thermal model to estimate $T_{\mathrm{atom}}$ from surrounding surfaces by 
\begin{equation}
T_{\mathrm{atom}}^{4}=\sum_{i}\Big(\frac{\Omega_{i}^{\mathrm{eff}}}{4\pi}\Big)T_{i}^{4},
\label{effectivetempeq}
\end{equation}	
where $T_{i}$ and $\Omega_{i}^{\mathrm{eff}}$ are the temperature and effective solid angle of a surface $i$, respectively. The temperature of the chamber surface is measured by an array of temperature sensors attached on the surface. The effective solid angle generally differs from the geometric solid angle, since the BBR photons reach atoms after being multiply scattered by chamber surfaces which are characterized by their emissivities and roughnesses. We estimate $\Omega_{i}^{\mathrm{eff}}$ by a Monte-Carlo tracking of the BBR photons with a chamber geometry shown in Figure \ref{geant4geometry}. Table \ref{effectivesolidangles} lists some examples of the calculated effective solid angles. The details of the Monte-Carlo simulation are given in our previous publication \cite{Kobayashi2022}, and thus we here describe two main differences from the previous work. (i) We divide the main chamber more finely to capture temperature inhomogeneity across the main chamber, while our previous work only treats the main chamber as one object and estimates its mean temperature from maximum and minimum values. Note that our previous simulation only estimates the contributions from hot sources such as an atomic oven and a heated window. (ii) The emissivity of the atomic oven nozzle (made of stainless-steel) is assigned as $\epsilon=0.1$, while the previous simulation assumes $\epsilon=1$. This is based on the fact that we find a small radiative temperature change ($\sim20$ mK) measured by the in-vacuum temperature sensor when blocking the atomic oven with a mechanical shutter. 

\begin{table}[h]
\caption{Effective solid angles calculated by the Monte-Carlo simulation.}  
%	\legend{\small{Only positive total $M$ values are shown. The table is symmetric for $M<0$.}}
	\label{effectivesolidangles}
	\begin{center} 
\begin{tabular}{lc}
\hline
Surface $i$  & $\Omega_{i}^{\mathrm{eff}}/(4\pi)$ \\
\hline
Main chamber A & 0.57 \\
Main chamber B & 0.15 \\
In-vacuum coil & 0.26 \\
Heated window & $3.7\times10^{-3}$ \\
Copper mount of the heated window & $2.1\times10^{-4}$\\ 
Tube between the heated window and the main chamber & $4.1\times10^{-3}$\\
Atomic oven nozzle & $7.8\times10^{-5}$\\
Radiation shield for the atomic oven & $3.0\times10^{-4}$\\
Zeeman slower tube & $6.2\times10^{-3}$ \\
\hline
\end{tabular}
\end{center}
\end{table}

\begin{figure}[h]
\includegraphics[scale=0.55]{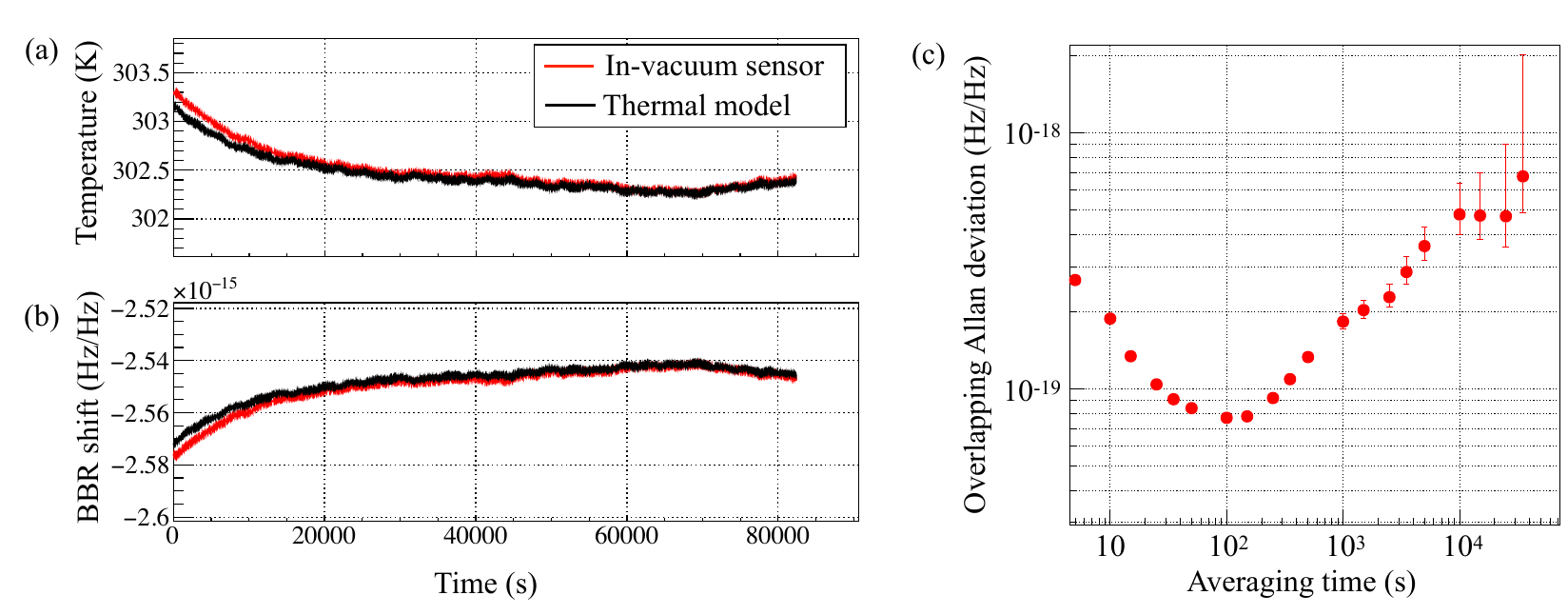}
\caption{Comparisons of (a) the radiative temperature at the position of the atoms and (b) the BBR shift between the in-vacuum sensor measurement and the thermal model estimate. (c) Overlapping Allan deviation calculated from the difference of the BBR shift between the two methods.}
\label{sensormodelcomparison}
\end{figure}

Figures \ref{sensormodelcomparison} (a) and (b) show comparisons of $T_{\mathrm{atom}}$ and $\Delta \nu_{\mathrm{BBR}}$, respectively, between the in-vacuum sensor measurement and the thermal model prediction. The long-term variation of $T_{\mathrm{atom}}$ observed in the two methods is mostly caused by that of the room temperature. The instability of the BBR shift difference between the two methods is shown in Fig.~\ref{sensormodelcomparison} (c), indicating the capability of the thermal model to track the radiative temperature variation for $\tau\lesssim10^{4}$ s. The mean difference of $\Delta \nu_{\mathrm{BBR}}$ between the two methods is $1.5(4.4)\times10^{-18}$ where the uncertainty is conservatively estimated by a peak-to-peak value divided by 2 of the sensor-model difference over several measurements conducted on different dates (not fully shown in Fig.~\ref{sensormodelcomparison}). Note that the agreement in the absolute temperature between the two methods is accidental, since the thermal model predicts $T_{\mathrm{atom}}$ with limited knowledge of $T_{i}$ and $\Omega_{i}^{\mathrm{eff}}$. 

\begin{figure}[h]
\includegraphics[scale=0.45]{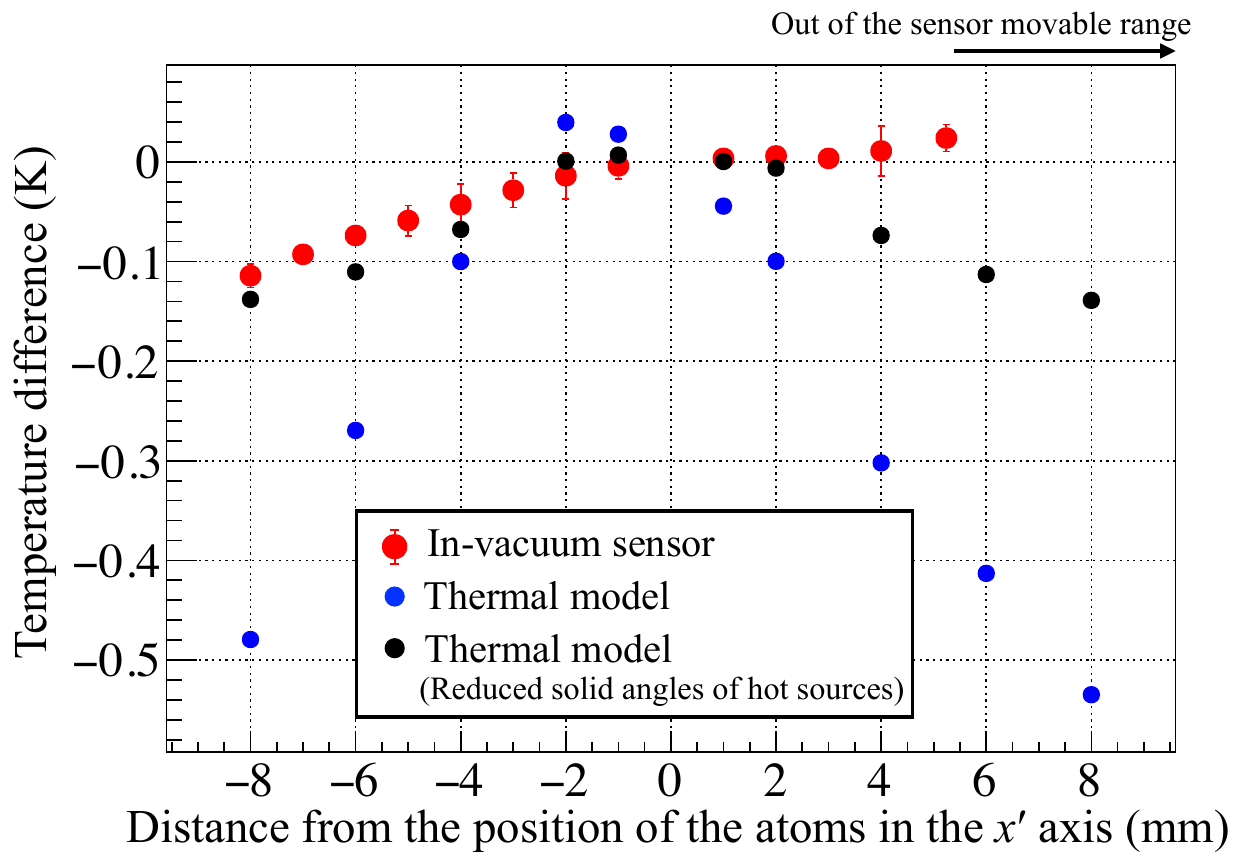}
\caption{Temperature difference from the temperature at the position of atoms as a function of the sensor position in the $x^{\prime}$ axis. The measured data (red point) are compared with the thermal model with nominal surface parameters (blue point) and the thermal model with a reduced effective solid angle of the atomic oven by a factor of 0.5 and that of heated window by a factor of 0.1 (black point).}
\label{positiondependence}
\end{figure}

To investigate the radiative temperature gradient near the position of atoms, we measure the temperature difference from the temperature at the atom position as a function of the sensor position in the $x^{\prime}$ axis (see Fig.~\ref{experimentalsetup}). The measurement result is shown by red points in Figure \ref{positiondependence}. Since the measurement is affected by the long-term fluctuation of the room temperature, we repeat the measurement at the atom position ($x^{\prime}=0$ mm) before and after the measurement at each sensor position ($x^{\prime}\ne0$ mm). The reproducibility of the temperature at the atom position is $\sim20$ mK in each measurement, which is included as an uncertainty in each data point in Fig.~\ref{positiondependence}. The measured temperature differences are compared with simulated differences by the thermal model. The thermal model (blue points in Fig.~\ref{positiondependence}) yields a larger temperature gradient caused by the radiation from the atomic oven and heated window. By intentionally reducing effective solid angles of the atomic oven and heated window, we find better agreement between the measurement and model (black points in Fig.~\ref{positiondependence}), implying the overestimation of the effective solid angles of the heat sources in the thermal model.

The temperature gradient causes an uncertainty of $T_{\mathrm{atom}}$ because of the position resolution of the sensor and ambiguity of the atom position in the $z$ direction (see Fig.~\ref{experimentalsetup}). From the measured temperature difference data (red points in Fig.~\ref{positiondependence}), we find a slope $\le11$ mK/mm near the atoms over a range between $x^{\prime}=\pm2$ mm. Taking account of the sensor length of 2.3 mm in the $x^{\prime}$ direction, an uncertainty of 11 $\mathrm{mK/mm}\times2.3$ mm$/2=13$ mK is assigned due to the position resolution of the sensor. Since the sensor does not move in the $y$, $z^{\prime}$, and $z$ directions, we employ the thermal model to estimate the temperature gradients in these directions. As discussed above, it is likely that the thermal model puts a conservative upper bound of the temperature gradient. The thermal model yields the absolute values of slopes $\le107$ mK/mm, $\le102$ mK/mm, and $\le73$ mK/mm from the atom position in the $y$, $z^{\prime}$, and $z$ directions, respectively. Since the sensor length is 2.1 mm (0.9 mm) in the $y$ ($z^{\prime}$) direction, an uncertainty of 107 $\mathrm{mK/mm}\times2.1$ mm$/2\sqrt{3}=65$ mK (102 $\mathrm{mK/mm}\times0.9$ mm$/2\sqrt{3}=27$ mK) is assigned assuming a rectangular distribution. From the intersection of the 399-nm and 578-nm guide lasers, we find that the atom position is $\le1.5$ mm above the sensor and estimate an uncertainty of $73$ $\mathrm{mK/mm}\times1.5$ $\mathrm{mm}/\sqrt{3}=63$ mK due to the atom position uncertainty in the $z$ direction. In total, the uncertainty due to the temperature gradient is estimated as 95 mK.

\begin{table}[h]
\caption{Uncertainty budget of the BBR shift.}  
%	\legend{\small{Only positive total $M$ values are shown. The table is symmetric for $M<0$.}}
	\label{budgetBBR}
	\begin{center} 
\begin{tabular}{lc}
\hline
Effect  & $\mathrm{Uncertainty}$ $(\times10^{-18}$ Hz/Hz) \\
\hline
Sensor calibration (3.0 mK) & 0.1 \\
Multimeter calibration (15 mK) & 0.5 \\
Self heating (1.5 mK) & 0.05 \\
Immersion error (32 mK) & 1.1\\
Temperature gradient (95 mK) & 3.2\\
Sensor-model difference & 4.4\\
Dynamic correction error & 4.1\\
Glass tube shielding (33 mK) & 1.1\\
Atomic coefficient & 0.7 \\
\hline
Total & 7.0\\
\hline
\end{tabular}
\end{center}
\end{table}

Table \ref{budgetBBR} shows the uncertainty budget of the BBR shift. The self heating effect of the sensor is measured by changing the excitation current from 10 $\mu$A to 3 mA by the ISOTEC thermometry bridge. The uncertainty due to the self heating at an operating current of 30 $\mu$A is estimated as 1.5 mK by dividing the observed self heating shift at this current by $\sqrt{3}$. The immersion error is evaluated by heating the flange to fix the sensor by 11 K above the temperature of the in-vacuum sensor. The observed increase of the sensor temperature is 50 mK, which is scaled down to 32 mK by a maximum temperature difference between the flange and sensor in the clock operation condition. The application of the temperature measured by the sensor in the dynamic corrections causes an error in the presence of temperature inhomogeneity in the chamber surface \cite{Nicholsonthesis,Nicholoson2015}. This is due to the fact that the sensor only yields a sold-angle weighted mean of $T_{i}^{4}$ (see Eq.~(\ref{effectivetempeq})) but not of $T_{i}^{6}$ or $T_{i}^{8}$ needed to calculate the dynamic corrections (see Eq.~(\ref{bbrshifteq})). To conservatively estimate the upper limit of this error with the thermal model, we recalculate $\Omega_{i}^{\mathrm{eff}}$ by assigning (i) a lower emissivity ($\epsilon = 0.05$) and specular surfaces for the stainless-steel chamber and (ii) higher emissivities for the heated window ($\epsilon=1$) and atomic oven ($\epsilon=0.3$) (see Ref.~\cite{Kobayashi2022} for the reason for choosing these surface parameters), which enhances the contribution from the hot sources. With resulting $\Omega_{i}^{\mathrm{eff}}$, the error is calculated as $4.1\times10^{-18}$ which is mostly due to the heated window. The insertion of the glass tube may change the radiative temperature at the atom position by partially shielding the radiations from the hot sources. We bound this effect to be $\le33$ mK with the thermal model. The contribution of the uncertainties of the atomic coefficients \cite{Sherman2012,Hassan2025} at our operating temperature is estimated as $7\times10^{-19}$. In total, the uncertainty of the BBR shift is $7.0\times10^{-18}$.

\subsection{Density shift}
\label{densityshiftresult}
\begin{figure}[h]
\includegraphics[scale=0.4]{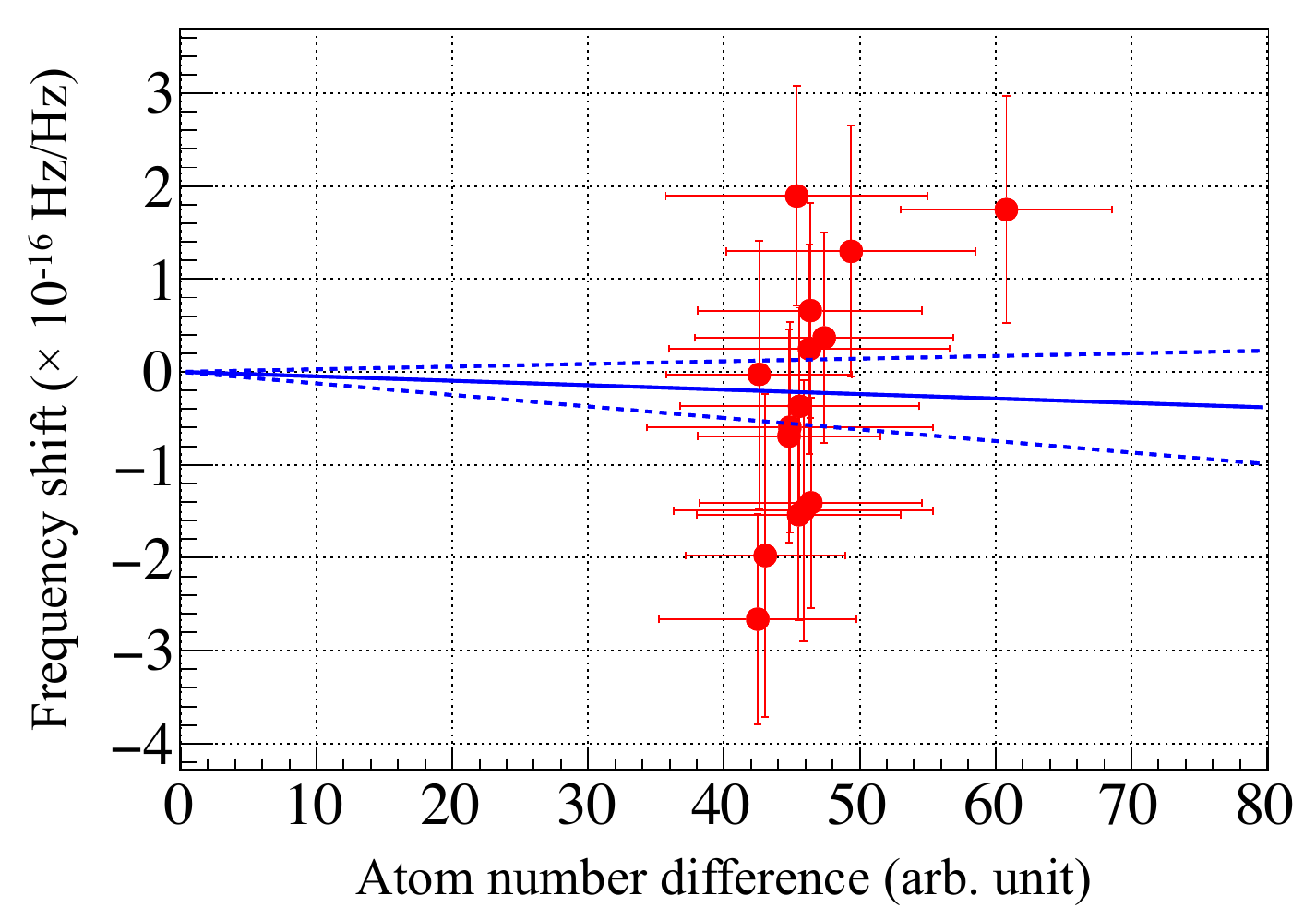}
\caption{Density shift as a function of the atom number difference. The solid (dashed) blue line indicates the linear fit (its uncertainty). The uncertainty is inflated by $\sqrt{\chi_{\mathrm{red}}^{2}}=1.1$.}
\label{densityfig}
\end{figure}

To measure the density shift, we carry out an interleaved measurement in which the number of the sideband-cooled atoms $N_{\mathrm{atom}}$ is varied between two values $N_{\mathrm{atom}}^{\mathrm{high}}$ and $N_{\mathrm{atom}}^{\mathrm{low}}$. Figure \ref{densityfig} shows measured shift values as a function of the atom number difference $N_{\mathrm{atom}}^{\mathrm{high}}-N_{\mathrm{atom}}^{\mathrm{low}}$ (arb. unit) at a trap depth of 511(29)$E_{\mathrm{r}}$, obtained in a total accumulation time of $\sim2$ days. The density shift $\Delta\nu_{\mathrm{D}}$ in the interleaved measurement is given by $\Delta\nu_{\mathrm{D}}=\alpha_{\mathrm{D}}(N_{\mathrm{atom}}^{\mathrm{high}}-N_{\mathrm{atom}}^{\mathrm{low}})$ where $\alpha_{\mathrm{D}}$ is the density shift coefficient. We deduce $\alpha_{\mathrm{D}}$ by a linear fit as shown in Fig.~\ref{densityfig}, and estimate the density shift at our operating atom number with a scaling $N_{\mathrm{atom}}U_{\mathrm{max}}^{5/4}$ \cite{Nicholsonthesis,Nicholoson2015}. For a typical operating condition of $N_{\mathrm{atom}}\sim80$ (arb. unit) and $U_{\mathrm{max}}\sim70E_{\mathrm{r}}$, the uncertainty of the density shift is $\sim6\times10^{-18}$.

\subsection{Other shifts}
The evaluations of the second-order Zeeman shift, probe light shift, servo error are same as those described in Refs.~\cite{Kobayashi2018,Kobayashi2020,Kobayashi2025}. 

The line pulling is estimated as $<2\times10^{-18}$ due to the fact that residual excitations of the $\pi$ and $\sigma$ transitions arising from imperfect spin polarization and misalignment of the clock laser polarization relative to the quantization axis are $<5$ $\%$. 

The uncertainty due to the AOM switching is estimated as $<1\times10^{-19}$ by counting the frequency of the phase-locked RF signal in the active stabilization of the optical path length including the AOM (see Section \ref{experimentalsetupsection}).

The DC Stark shift arising from the glass tube to support the in-vacuum temperature sensor is estimated by (i) bounding the DC electric potential of the glass tube as $<100$ V with an electrostatic sensor (Keyence SK-050) before installing the sensor into vacuum and (ii) performing a finite element analysis to estimate the electric field at the atom position. We bound the DC Stark shift as $<3\times10^{-18}$ using the reported static polarizability \cite{Sherman2012}.

The background gas shift due to collisions with the background gas and atomic beam is estimated as $-2.7(8)\times10^{-18}$ using a measured trap lifetime of 5.9(1.4) s and a theoretical coefficient of $-1.6(3)\times10^{-17}$ s \cite{Pizzocaro2020}. The theoretical coefficient includes the effects of the collisions between (i) Yb and H$_{2}$ and (ii) Yb and Yb, and is consistent with an experimental coefficient $-1.64(12)\times10^{-17}$ s \cite{McGrew2018} for the collision between Yb and H$_{2}$. To separate the contribution from the background gas and that from the atomic beam, we temporarily actuate the mechanical shutter for the atomic beam and measure the lifetime. The resulting lifetime is 5.3(1.4) s in agreement with the above lifetime in the presence of the atomic beam, implying the domination of the contribution from the background gas. Note that we normally do not actuate the shutter, since it is not designed for long-term use. 

\section{Nearly continuous operation for contributing to TAI}
\label{nearlycontinuoussection}
NMIJ-Yb1 is linked to TAI by the relationship \cite{Kobayashi2025}
\begin{eqnarray}
y(\mathrm{TAI-NMIJ\mathchar`-Yb1})&=&y(\mathrm{TAI-UTC(NMIJ)})\nonumber\\
&&+y(\mathrm{UTC(NMIJ)-HM})\nonumber\\
&&+y(\mathrm{HM-NMIJ\mathchar`-Yb1}),
\end{eqnarray}
where $y(\mathrm{A-B})$ denotes the fractional frequency difference between A and B, HM the hydrogen maser, and UTC(NMIJ) Coordinated Universal Time of NMIJ. 
To contribute to TAI, we measure $y(\mathrm{HM-NMIJ\mathchar`-Yb1})$ using the frequency comb (see Section \ref{experimentalsetupsection}) and calculate $y(\mathrm{UTC(NMIJ)-HM})$ from time difference data of 1 pulse-per-second (1 PPS) signals between HM and UTC(NMIJ). $y(\mathrm{TAI-UTC(NMIJ)})$ is derived from data based on global navigation satellite systems (GNSSs), calculated by Bureau International des Poids et Mesures (BIPM) \cite{circulart}. Since $y(\mathrm{TAI-UTC(NMIJ)})$ is provided for an averaging time of consecutive days that are multiples of 5, we estimate the mean values of $y(\mathrm{HM-NMIJ\mathchar`-Yb1})$ and $y(\mathrm{UTC(NMIJ)-HM})$ over the same averaging time. For this, the extrapolation of $y(\mathrm{HM-NMIJ\mathchar`-Yb1})$ is carried out due to the dead time of NMIJ-Yb1. The uncertainty arising from the extrapolation is estimated based on the frequency stability model of the hydrogen maser \cite{Kobayashi2024}, which is reduced to a negligible level by operating NMIJ-Yb1 with a high uptime. 

\begin{figure}[h]
\includegraphics[scale=0.6]{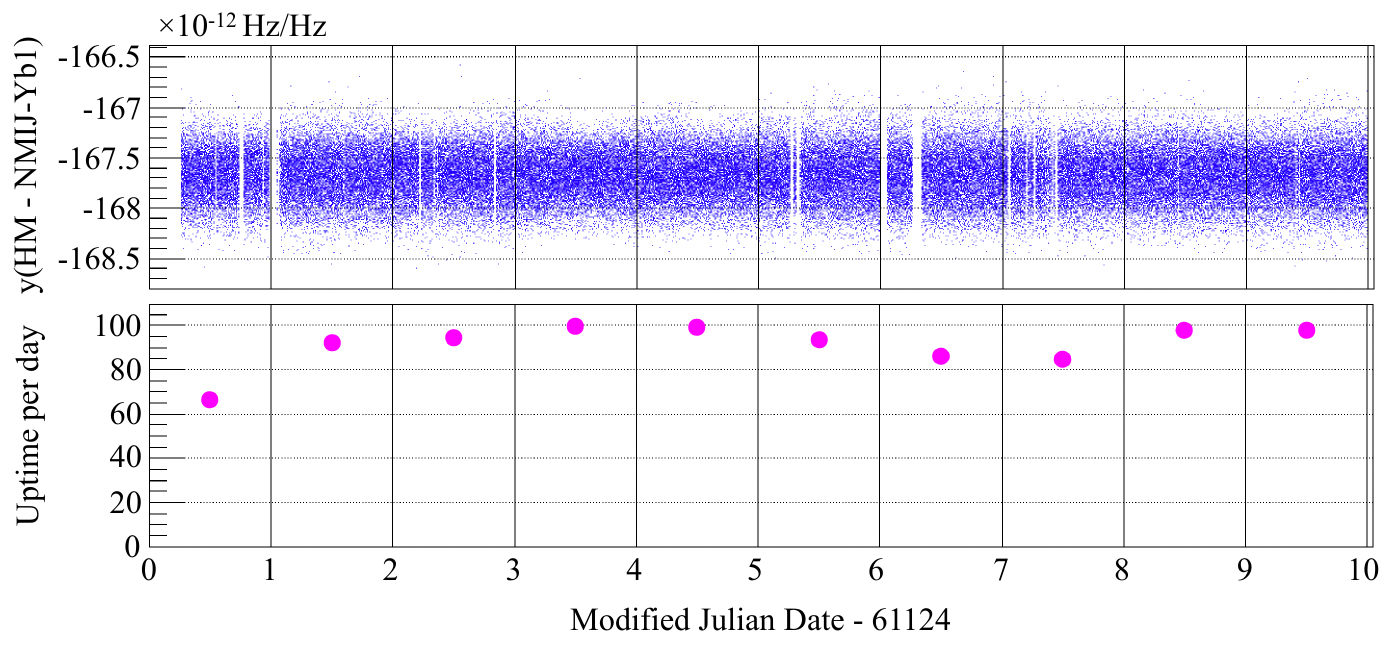}
\caption{Fractional frequency difference between the hydrogen maser (HM) and NMIJ-Yb1 averaged over 6.8 s and uptime of NMIJ-Yb1 per day. }
\label{operationfigure}
\end{figure}

Since our primary motivation for improving the systematic uncertainty of NMIJ-Yb1 is to improve the uncertainty of the TAI calibration, it is essential to keep the capability of the operation with a high uptime. To demonstrate the robustness of NMIJ-Yb1 after the updates, we operate NMIJ-Yb1 for 10 days as shown in Figure \ref{operationfigure}. The uptime for the 10-day period is 91.3 $\%$, which is sufficiently high to reduce the extrapolation uncertainty to $<1\times10^{-16}$. 

During the 10-day period from March 25 (Modified Julian Date 61124) to April 4 in 2026, $y(\mathrm{TAI}-$NMIJ-Yb1$)$ is evaluated as $2.2(3.9)\times10^{-16}$ using the recommended frequency $518\,295\,836\,590\,863.632$ Hz of $^{171}$Yb \cite{recommendedvalue}. Our evaluation result is in good agreement with weighted mean values (i) $1.5(1.0)\times10^{-16}$ and (ii) $1.2(1.0)\times10^{-16}$ of TAI evaluations $y(\mathrm{TAI}-$PSFS$)$ by primary and secondary frequency standards (PSFSs) (i) from February 28 to March 30 and (ii) from March 30 to April 29, respectively \cite{circulart}\footnote{The evaluation period of TAI is divided by convention \cite{circulart}.}. Our uncertainty $3.9\times10^{-16}$ arises from the effects summarized in Table \ref{taicalibrationbudge}. Note that the uncertainty of the recommended frequency ($u_{\mathrm{SecRep}}=1.7\times10^{-16}$) is not included, and that the statistical uncertainty $u_{\mathrm{A/Lab}}$ of the local link is mostly due to the measurement uncertainty of the 1 PPS signals, not the extrapolation uncertainty. We here improve the systematic uncertainty $u_{\mathrm{B/Lab}}$ of the local link by employing a coaxial cable with a smaller temperature dependence \cite{Wada2022}, compared with our previous uncertainty $1\times10^{-16}$ which has also been a non-negligible factor for the criterion requiring $\lesssim2\times10^{-16}$ uncertainty. 

\begin{table}[h]
\caption{Uncertainty budget of the TAI calibration by NMIJ-Yb1 during the 10-day period from March 25 to April 4 in 2026. Many details of the TAI calibration are described elsewhere \cite{Kobayashi2020,Kobayashi2025}, and so we only provide the values here.}  
%	\legend{\small{Only positive total $M$ values are shown. The table is symmetric for $M<0$.}}
	\label{taicalibrationbudge}
	\begin{center} 
\begin{tabular}{lc}
\hline
Effect  & $\mathrm{Uncertainty}$ $(\times10^{-16}$ Hz/Hz) \\
\hline
NMIJ-Yb1 (statistics) $u_{\mathrm{A}}$ & 0.11 \\
NMIJ-Yb1 (systematics) $u_{\mathrm{B}}$ & 0.27 \\
Local link to UTC(NMIJ) (statistics) $u_{\mathrm{A/Lab}}$ &  1.6\\
Local link to UTC(NMIJ) (systematics) $u_{\mathrm{B/Lab}}$ & 0.54\\
GNSS link to TAI $u_{\mathrm{l/Tai}}$ & 3.5\\
\hline
Total & 3.9\\
\hline
\end{tabular}
\end{center}
\end{table}

\section{Discussions and conclusion}
\label{discussionsection}
To investigate the model-dependent bias of the lattice light shift, we perform an additional analysis with a different light shift model \cite{Katori2015,Ushijima2018} based on the harmonic oscillator approximation. This model yields the lattice light shift by
\begin{eqnarray}
\Delta \nu_{\mathrm{LS}}&=&(\alpha_{E1}\Delta \nu - \alpha_{M1E2})\Big(\Braket{n_{z}}+\frac{1}{2}\Big)\Big(U_{\mathrm{e}}/E_\mathrm{r}\Big)^{1/2}\nonumber\\
&&-\Big\{\alpha_{E1}\Delta \nu + \frac{3}{4}\beta\Big(2\Braket{n_{z}}^{2}+2\Braket{n_{z}}+1\Big)\Big\}\Big(U_{\mathrm{e}}/E_\mathrm{r}\Big )\nonumber\\
&&+\beta(2\Braket{n_{z}}+1)\Big(U_{\mathrm{e}}/E_{\mathrm{r}}\Big)^{3/2}\nonumber\\
&&-\beta\Big(U_{\mathrm{e}}/E_{\mathrm{r}}\Big)^2,
\label{lightshifteq}
\end{eqnarray}
where $\Braket{n_{z}}\lesssim0.3$ is the mean axial quantum number calculated by the ratio of the red and blue sidebands, and $U_{\mathrm{e}}^{j}=(1+j k_{\mathrm{B}} T_{\mathrm{r}}/U_{\mathrm{max}})^{-1}U_{\mathrm{max}}^{j}$ is the effective trap depth to take account of the depth reduction due to the radial motion, where $j$ denotes the power series exponent for each term in Eq.~(\ref{lightshifteq}) \cite{Ushijima2018,Beloy2020}. As summarized in Table \ref{modelcomparison}, $\nu_{E1}$ and $\Delta\nu_{\mathrm{LS}}$ derived from the harmonic oscillator model agree with those from the Born-Oppenheimer + WKB model within the 1 $\sigma$ uncertainties. Note that the uncertainties of $\nu_{E1}$ and $\Delta\nu_{\mathrm{LS}}$ are mainly limited by the statistical uncertainty of the interleaved measurement, and that we find a model-dependent bias of $4\times10^{-18}$ in $\Delta\nu_{\mathrm{LS}}$.

We also compare $T_{\mathrm{r}}$ derived from the fit of the sideband spectrum based on the Born-Oppenheimer approximation (see Eq.~(\ref{sidebandfitfunc})) with $T_{\mathrm{r}}$ from a conventional fit based on a harmonic oscillator potential \cite{Blatt2009}. In both cases, $T_{\mathrm{r}}$ is derived assuming a Boltzmann distribution, and thus is dependent on the model of the radial potential that determines the motional energies of atoms. As shown in Table \ref{modelcomparison}, we find disagreement of $T_{\mathrm{r}}$ between the two models. Previous studies \cite{Goti2025} have also reported the similar discrepancy, where the harmonic oscillator fit yields higher radial temperatures. At the present experimental uncertainty level, however, $\nu_{E1}$ and $\Delta \nu_{\mathrm{LS}}$ derived from the harmonic oscillator fit are still in agreement with those from the Born-Oppenheimer fit.  

\begin{table}[h]
\caption{Comparisons of the lattice light shift models. In each model, $\nu_{E1}$ is derived from the data taken in the interleaved measurement. The obtained $\nu_{E1}$ is utilized to calculate $\Delta\nu$ in the operational condition for the TAI calibration. BO: Born-Oppenheimer, HO: Harmonic oscillator.} 
%	\legend{\small{Only positive total $M$ values are shown. The table is symmetric for $M<0$.}}
	\label{modelcomparison}
	\begin{center} 
\begin{tabular}{lccc}
\hline
 & BO + WKB & HO & HO \\
 & BO sideband fit & BO sideband fit & HO sideband fit \\
\hline
Interleaved measurement & & &\\
$\nu_{E1}$ (MHz) & $394\,798\,261.0(5.8)$ & $394\,798\,258.7(5.7)$ & $394\,798\,268.4(7.8)$\\
\hline
Operational condition & & & \\
$T_{r}$ ($\mu$K) & 0.73(15) &  0.73(15) &  3.9(8)\\
$\Delta\nu_{\mathrm{LS}}$ ($\times10^{-17}$ Hz/Hz)& $4.4(1.6)$& 4.0(1.6)&  4.7(1.6) \\
\hline
\end{tabular}
\end{center}
\end{table}

By extending the evaluation period of TAI, the limiting statistical uncertainties $u_{\mathrm{A/Lab}}$ and $u_{\mathrm{l/Tai}}$ in Table \ref{taicalibrationbudge} are reduced. For example, we have operated NMIJ-Yb1 with an uptime of 86.6 $\%$ for 30 days in June 2025, which yields $u_{\mathrm{A/Lab}}=0.7\times10^{-16}$ and $u_{\mathrm{l/Tai}}=1.7\times10^{-16}$ \cite{circulart}\footnote{For 30-day period, $u_{\mathrm{l/Tai}}$ can reach as low as $1.3\times10^{-16}$ depending on the quality of the GNSS data evaluated by BIPM \cite{circulart}}. These statistical uncertainties would result in the total calibration uncertainty of $1.9\times10^{-16}$ (see Table \ref{taicalibrationbudge}) that meets the criterion. For comparison, with our previous systematic uncertainties $u_{\mathrm{B}}\sim1\times10^{-16}$ and $u_{\mathrm{B/Lab}}\sim1\times10^{-16}$, the total calibration uncertainty in June 2025 is $2.3\times10^{-16}$ which marginally meets the criterion.

To validate the systematic uncertainty at the low $10^{-17}$, we need to compare NMIJ-Yb1 with another optical clock developed at NMIJ, optical clocks at different laboratories via fiber links \cite{Hong2009,Akatsuka2020}, or transportable optical clocks \cite{Ohmae2021,Nosske2025,Bothwell2025optlett}. In the near future plan, NMIJ-Yb1 will be compared with our Sr optical lattice clock \cite{Akamatsu2014,Hisai2021} after reducing its systematic uncertainty. This local frequency ratio measurement of the different transitions also serves as an independent measurement for international comparisons without complications inherent in fiber links or transportable optical clocks. To measure the frequency ratio as precisely as possible, we plan to further improve the systematic uncertainty of NMIJ-Yb1 to $\lesssim1\times10^{-17}$ limited by the uncertainty of the BBR shift ($7.0\times10^{-18}$). This can be done by choosing the experimental parameters that reduce the uncertainties of the limiting factors such as the lattice light shift and servo error. To reduce the uncertainty of the lattice light shift, we can set the operating trap depth to $<70E_{\mathrm{r}}$, while the nominal depth $\sim70E_{\mathrm{r}}$ is chosen to ensure a sufficient margin in the number of atoms for long-term operation. Regarding the servo error, the uncertainty is limited by the statistical uncertainty calculated by the data taken for 10 days (see Section \ref{nearlycontinuoussection}), and can be reduced to $<1\times10^{-17}$ for an extended operation period \cite{circulart}. Furthermore, the linewidth of the carrier transition is expected to be reduced with the crystalline mirror cavity \cite{Nishiyama}, which also contributes to improve the servo error uncertainty.

In conclusion, we have improved the systematic uncertainty of NMIJ-Yb1 to $2.6\times10^{-17}$. We have also demonstrated the nearly continuous operation of NMIJ-Yb1 for 10 days, and found the agreement of the TAI calibration result with those by other PSFSs at the mid $10^{-16}$ level. NMIJ-Yb1 will contribute to achieve the criterion for the TAI contribution towards the redefinition of the SI second.

\section*{Acknowledgments}
We are grateful to K Hosaka for discussion and assistance to operate NMIJ-Yb1, H Inaba and M Wada for development of the frequency comb and RF synthesis system, T Tanabe, A Iwasa and Y Fujii for maintaining UTC(NMIJ), and Y Kawamura and H Ogura for discussion of the temperature sensor calibration. We are indebted to national metrology institutes for operating the primary and secondary frequency standards and the BIPM time department for making the evaluation results available. T K acknowledges neutral atom optical clocks group at National Institute of Standards and Technology (NIST) for hosting the sabbatical stay, which inspires T K to improve NMIJ-Yb1. This work was supported by Japan Society for the Promotion of Science (JSPS) KAKENHI Grant Number 22H01241, 22K04942 and 26H02059, JST-Mirai Program Grant Number JPMJMI18A1, and the JST Moonshot R $\&$ D Program Grant Number JPMJMS2268, Japan. 

\section*{Appendix A. Reduction factors of the trap depth based on the Born-Oppenheimer + WKB model}
The reduction factors $X(n_{z},U_{\mathrm{max}},T_{\mathrm{r}})$, $Y(n_{z},U_{\mathrm{max}},T_{\mathrm{r}})$, and $Z(n_{z},U_{\mathrm{max}},T_{\mathrm{r}})$ in Eq.~(\ref{bolightshifteq}) represent the ensemble averages of the expectation values of the spatial portion of the trap potential $e^{-2\rho^{2}/w^{2}}\cos^{2}(kz)$, $e^{-2\rho^{2}/w^{2}}\sin^{2}(kz)$, and $e^{-4\rho^{2}/w^{2}}\cos^{4}(kz)$, respectively, where $k=2\pi\nu_{\mathrm{L}}/c$. To calculate the reduction factors, we employ the Born-Oppenheimer + WKB model \cite{Beloy2020} because of its nonperturbative approach to include the axial-radial coupling and the anharmonicity in the trap potential. The Born-Oppenheimer + WKB model yields 
\begin{equation}
X(n_{z},U_{\mathrm{max}},T_{\mathrm{r}})=\frac{\int^{\rho_{n_{z}}^\mathrm{max}}_{0}x_{n_{z}}(\rho)\rho(e^{-U_{n_{z}}(\rho)/k_{\mathrm{B}}T_{\mathrm{r}}}-1)d\rho}{\int^{\rho_{n_{z}}^\mathrm{max}}_{0}\rho(e^{-U_{n_{z}}(\rho)/k_{\mathrm{B}}T_{\mathrm{r}}}-1)d\rho},
\end{equation}
\begin{equation}
Y(n_{z},U_{\mathrm{max}},T_{\mathrm{r}})=\frac{\int^{\rho_{n_{z}}^\mathrm{max}}_{0}y_{n_{z}}(\rho)\rho(e^{-U_{n_{z}}(\rho)/k_{\mathrm{B}}T_{\mathrm{r}}}-1)d\rho}{\int^{\rho_{n_{z}}^\mathrm{max}}_{0}\rho(e^{-U_{n_{z}}(\rho)/k_{\mathrm{B}}T_{\mathrm{r}}}-1)d\rho},
\end{equation}
\begin{equation}
Z(n_{z},U_{\mathrm{max}},T_{\mathrm{r}})=\frac{\int^{\rho_{n_{z}}^\mathrm{max}}_{0}z_{n_{z}}(\rho)\rho(e^{-U_{n_{z}}(\rho)/k_{\mathrm{B}}T_{\mathrm{r}}}-1)d\rho}{\int^{\rho_{n_{z}}^\mathrm{max}}_{0}\rho(e^{-U_{n_{z}}(\rho)/k_{\mathrm{B}}T_{\mathrm{r}}}-1)d\rho},
\end{equation}
where $U_{n_{z}}(\rho)$ is the trap potential and $\rho^{\mathrm{max}}_{n_z}$ is the radial position at which $U_{n_{z}}(\rho)/E_{\mathrm{r}}$ is zero.  $U_{n_{z}}(\rho)$ is given by
\begin{equation}
U_{n_{z}}(\rho)=E_{\mathrm{r}}\Big[b_{n_{z}+1}\Big(D(\rho)/4\Big)-D(\rho)/2\Big],
\label{mathiefunction}
\end{equation}
where $b_{r}(q)$ is the characteristic value for the odd Mathieu function and $D(\rho)=(U_{\mathrm{max}}/E_{\mathrm{r}})e^{-2\rho^{2}/w^{2}}$ is the radial potential for the lattice beam waist $w$. Note that $U_{n_{z}}(\rho)$ is also used in the sideband fit in Eq.~(\ref{sidebandfitfunc}).
$x_{n_{z}}(\rho)$, $y_{n_{z}}(\rho)$, and $z_{n_{z}}(\rho)$ are given by
\begin{equation}
x_{n_{z}}(\rho)=e^{-2\rho^{2}/w^{2}}\int^{+\pi/2k}_{-\pi/2k}|\mathcal{Z}_{n_{z}}(\rho, z)|^{2}\cos^{2}(kz)dz,
\end{equation}
\begin{equation}
y_{n_{z}}(\rho)=e^{-2\rho^{2}/w^{2}}\int^{+\pi/2k}_{-\pi/2k}|\mathcal{Z}_{n_{z}}(\rho, z)|^{2}\sin^{2}(kz)dz,
\end{equation}
\begin{equation}
z_{n_{z}}(\rho)=e^{-4\rho^{2}/w^{2}}\int^{+\pi/2k}_{-\pi/2k}|\mathcal{Z}_{n_{z}}(\rho, z)|^{2}\cos^{4}(kz)dz,
\end{equation}
where $\mathcal{Z}_{n_{z}}(\rho, z)$ is the eigenfunction to yield the eigenvalue $U_{n_{z}}(\rho)$ in the treatment of the Born-Oppenheimer approximation. $\mathcal{Z}_{n_{z}}(\rho, z)$ is given by
\begin{equation}
\mathcal{Z}_{n_{z}}(\rho, z)=\sqrt{\frac{2k}{\pi}}se_{n_z+1}\Big(kz+\pi/2,D(\rho)/4 \Big),
\end{equation}
where $se_{r}(z,q)$ is the odd Mathieu function.


\section*{Reference}
\begin{thebibliography}{1}
\bibitem{Ushijima2015} Ushijima I, Takamoto M, Das M, Ohkubo T, and Katori H 2015 Cryogenic optical lattice clocks $Nat.$ $Photon.$ $\bf{6}$ 185-189
\bibitem{Hunteman2016} Huntemann N, Sanner C, Lipphardt B, Tamm C, and Peik E 2016 Single-ion atomic clock with $3\times10^{-18}$ systematic uncertainty $Phys.$ $Rev.$ $Lett.$ $\bf{116}$ 063001 
\bibitem{McGrew2018} McGrew W F, Zhang X, Fasano R J, Sch\"affer S A, Beloy K, Nicolodi D, Brown R C, Hinkley N, Milani G, Schioppo M, Yoon T H, and Ludlow A D 2018 Atomic clock performance enabling geodesy below the centimetre level $Nature$ $\bf{564}$ 87-90
\bibitem{Brewer2019} Brewer S M, Chen J-S, Hankin A M, Clements E R, Chou C W, Wineland D J, Hume D B, and Leibrandt D R 2019 $^{27}$Al$^{+}$ Quantum-Logic Clock with a Systematic Uncertainty below $10^{-18}$ $Phys.$ $Rev.$ $Lett.$ $\bf{123}$ 033201
\bibitem{Bothwell2019} Bothwell T, Kedar D, Oelker E, Robinson J M, Bromley S L, Tew W L, Ye J, and Kennedy C J 2019 JILA SrI optical lattice clock with uncertainty of $2.0\times10^{-18}$ $Metrologia$ $\bf{56}$ 066004.
\bibitem{Tofful2024} Tofful A, Baynham C F A, Curtis E A, Parsons A O, Robertson B I, Schioppo M, Tunesi J, Margolis H S, Hendricks R J, Whale J, Thompson R C, and Godun R M 2024 $^{171}$Yb$^{+}$ optical clock with $2.2\times10^{-18}$ systematic uncertainty and absolute frequency measurements $Metrologia$ $\bf{61}$ 045001.
\bibitem{Li2024} Li J, Cui X-Y, Jia Z-P, Kong D-Q, Yu H-W, Zhu X-Q, Liu X-Y, Wang D-Z, Zhang X, Huang X-Y, Zhu M-Y, Yang Y-M, Hu Y, Liu X-P, Zhai X-M, Liu P, Jiang X, Xu P, Dai H-N, Chen Y-A and Pan J-W 2024 A strontium lattice clock with both stability and uncertainty below $5\times10^{-18}$ $Metrologia$ $\bf{61}$ 015006.
\bibitem{Aeppli2024} Aeppli A, Kim K, Warfield W, Safronova M S, and Ye J 2024 Clock with $8\times10^{-19}$ Systematic Uncertainty $Phys.$ $Rev.$ $Lett.$ $\bf{133}$ 023401
\bibitem{Lu2025} Lu X-T, Guo F, Liu Y-Y, Cao J, Li J-A, Xia J-J, Xu Q-F, Lu B-Q, Wang Y-B and Chang H 2025 NTSC SrII optical lattice clock with uncertainty of $2\times10^{-18}$ $Metrologia$ $\bf{62}$ 035007
\bibitem{Nosske2025} Nosske I, Vishwakarma C, L\"ucke T, Rahm J, Poudel N, Weyers S, Benkler E, D\"orscher S and Lisdat C 2025 Transportable strontium lattice clock with $4\times10^{-19}$ blackbody radiation shift uncertainty $Quantum$ $Sci.$ $Technol.$ $\bf{10}$ 045076
\bibitem{Lindvall2025pra} Lindvall T, Fordell T, Hanhij\"arvi K J, Dole\v{z}al M, Rahm J, Weyers S, and Wallin A E 2025 $^{88}$Sr$^{+}$ optical clock with $7.9\times10^{-19}$ systematic uncertainty and measurement of its absolute frequency with $9.8\times10^{-17}$ uncertainty $Phys.$ $Rev.$ $Appl.$ $\bf{24}$ 044082
\bibitem{Jia2026} Jia Z-P, Li J, Kong D-Q, Zhang X, Yu H-W, Liu X-Y, Zhang Y-C, Wang Y-B, Zhu X-Q, Zhang J-H, Zhu M-Y, Feng P-J, Cui X-Y, Xu P, Jiang X, Liu X-P, Liu P, Dai H-N, Chen Y-A, and Pan J-W 2026 Improved systematic evaluation of a strontium optical clock with uncertainty below $1\times10^{-18}$ $Metrologia$ $\bf{63}$ 025002
\bibitem{Zhang2026} Zhang T, Jin T, Qi Q, Lei S, Xia Y, Zhang J, Chang H, Feng S, Liu X, Wang J, Zhang R, Xu Z, Tang Z and Xu X 2026 $^{171}$Yb optical lattice clock with uncertainty below $5\times10^{-18}$ $Metrologia$ $\bf{63}$ 025004
\bibitem{ZhangB2026} Zhang B, Ma Z, Huang Y, Han H, Hu R, Wang Y, Zhang H, Tang L, Shi T, Guan H, and Gao K 2026 Liquid-Nitrogen-Cooled $^{40}$Ca$^{+}$ Ion Optical Clock with a Systematic Uncertainty of $4.4 \times 10^{-19}$ $Phys.$ $Rev.$ $Lett.$ $\bf{136}$ 053202
\bibitem{Zhu2026} Zhu Q, Shi J, Zhang Y,  Xu X, Wang B, Xiong D, Xiong Z, Fang P, Chen Q, He L, and Lyu B 2026 Ytterbium lattice clock with uncertainty of $1.1\times10^{-18}$ and instability of low $10^{-19}$ arXiv:2606.10514
\bibitem{Takano2016} Takano T, Takamoto M, Ushijima I, Ohmae N, Akatsuka T, Yamaguchi A, Kuroishi, Munekane H, Miyahara B, and Katori H 2016 Geopotential measurements with synchronously linked optical lattice clocks $Nat.$ $Photon.$ $\bf{10}$ 662-666
\bibitem{Grotti2018} Grotti J $et$ $al$. 2018 Geodesy and metrology with a transportable optical clock $Nat.$ $Phys.$ $\bf{14}$ 437-441
\bibitem{Grotti2024} Grotti J, Nosske I, Koller S B, Herbers S, Denker H, Timmen L, Vishnyakova G, Grosche G, Waterholter T, Kuhl A, Koke S, Benkler E, Giunta M, Maisenbacher L, Matveev A, D\"orscher S, Schwarz R, Al-Masoudi A, H\"ansch T W, Udem T, Holzwarth R, and Lisdat C 2024 Long-distance chronometric leveling with a portable optical clock $Phys.$ $Rev.$ $Appl.$ $\bf{21}$ L061001
\bibitem{Liu2024} Liu D, Wu L, Xiong C, and Bao L 2024 Geopotential Difference Measurement Using Two Transportable Optical Clocks' Frequency Comparisons $Remote$ $Sens.$ $\bf{16}$ 2462
\bibitem{Wcislo2018} Wcis\l o P $et$ $al.$ 2018 New bounds on dark matter coupling from a global network of optical atomic clocks $Sci.$ $Adv.$ $\bf{4}$, eaau4869
\bibitem{Takamoto2020} Takamoto M, Ushijima I, Ohmae N, Yahagi T, Kokado K, Shinkai H, and Katori H 2020 Test of General Relativity by a Pair of Transportable Optical Lattice Clocks $Nat$. $Photon$. $\bf{14}$ 411-415
\bibitem{Kennedy2020} Kennedy C J, Oelker E, Robinson J M, Bothwell T, Kedar D, Milner W R, Marit G E, Derevianko A, and Ye J 2020 Precision metrology meets cosmology: Improved constraints on ultralight dark matter from atom-cavity frequency comparisons $Phys.$ $Rev.$ $Lett.$ $\bf{125}$ 201302
\bibitem{Lange2021} Lange R, Huntemann N, Rahm J M, Sanner C, Shao H, Lipphardt B, Tamm C, Weyers S, and Peik E 2021 Improved limits for violations of local position invariance from atomic clock comparisons $Phys.$ $Rev.$ $Lett.$ $\bf{126}$ 011102
\bibitem{Kobayashi2022} Kobayashi T, Takamizawa A, Akamatsu D, Kawasaki A, Nishiyama A, Hosaka K, Hisai Y, Wada M, Inaba H, Tanabe T, and Yasuda M 2022 Search for ultralight dark matter from long-term frequency comparisons of optical and microwave atomic clocks $Phys.$ $Rev.$ $Lett.$ $\bf{129}$ 241301
\bibitem{Filzinger2023} Filzinger M, D\"orscher S, Lange R, Klose R J, Steinel M, Benkler E, Peik E, Lisdat C, and Hunteman N 2023 Improved limits on the coupling of ultralight bosonic dark matter to photons from optical atomic clock comparisons $Phys.$ $Rev.$ $Lett.$ $\bf{130}$ 253001
\bibitem{Kawasaki2025} Kawasaki A 2025 Quantum sensing using atomic clocks for nuclear and particle physics $Appl.$ $Phys.$ $Rev.$ $\bf{12}$ 041331
\bibitem{Hong2016} Hong F-L 2016 Optical frequency standards for time and length applications $Meas.$ $Sci.$ $Technol.$ $\bf{28}$ 012002
\bibitem{Dimarcq2024} Dimarcq N, Gertsvolf M, Mileti G, Bize S, Oates C W, Peik E, Calonico D, Ido T, Tavella P, and Meynadier F $et$ $al.$ 2024 Roadmap towards the redefinition of the second $Metrologia$ $\bf{61}$ 012001
\bibitem{CGPM2026} General Conference on Weights and Measures (CGPM) 2026 $Draft$ $Resolution$ B $of$ $its$ $28th$ $Meeting$
\bibitem{Sanner2019} Sanner C, Huntemann N, Lange R, Tamm C, Peik E, Safronova M S and Porsev S G 2019 Optical clock comparison for Lorentz symmetry testing Nature $\bf{567}$ 204
\bibitem{Hausser2025} Hausser H N, Keller J, Nordmann T, Bhatt N M, Kiethe J, Liu H, Richter I M, von Boehn M, Rahm J, Weyers S, Benkler E, Lipphardt B, D\"orscher S, Stahl K, Klose J, Lisdat C, Filzinger M, Huntemann N, Peik E, and Mehlst\"aubler T E 2025 $^{115}$In$^{+}$ - $^{172}$Yb$^{+}$ Coulomb Crystal Clock with $2.5 \times 10^{-18}$ Systematic Uncertainty $Phys.$ $Rev.$ $Lett.$ $\bf{134}$ 023201 
\bibitem{Aeppli2025} Aeppli A $et$ $al.$ 2026 Atomic clock frequency ratios with fractional uncertainty $\le 3.2 \times 10^{-18}$ $Phys.$ $Rev.$ $Lett.$ $\bf{137}$ 033201
\bibitem{Lindvall2025} Lindvall T $et$ $al.$ 2025 Coordinated international comparisons between optical clocks connected via fiber and satellite links $Optica$ $\bf{12}$ 843-852 
\bibitem{Amy-Klein2024} Amy-Klein A $et$ $al.$ 2024 International comparison of optical frequencies with transportable optical lattice clocks arXiv:2410.22973
\bibitem{Dawel2026} Dawel F, Kramer J, Drapier D, Pelzer L, Dietze K, Ali M A, Hild M, Barb\'e V, King S A, Klose J, Stahl K, Rahm J, Poudel N, D\"orscher S, Weyers S, Benkler E, Lisdat C, and Schmidt P O 2026 An Al$^{+}$ clock with $1.6\times10^{-18}$ systematic uncertainty and its frequency ratios arXiv:2606.23151
\bibitem{Pizzocaro2026} Pizzocaro M $et$ $al.$ 2026 International optical clock comparison using the European optical fiber network $Phys.$ $Rev.$ $Res.$ $\bf{8}$ 033250
\bibitem{circulart} BIPM Circular T (https://www. bipm.org/en/time-ftp)
\bibitem{Kobayashi2020} Kobayashi T, Akamatsu D, Hosaka K, Hisai Y, Wada M, Inaba H, Suzuyama T, Hong F-L, and Yasuda M 2020 Demonstration of the nearly continuous operation of an $^{171}$Yb optical lattice clock for half a year $Metrologia$ $\bf{57}$ 065021
\bibitem{Kobayashi2018} Kobayashi T, Akamatsu D, Hisai Y, Tanabe T, Inaba H, Suzuyama T, Hong F-L, Hosaka K, and Yasuda M 2018 Uncertainty Evaluation of an $^{171}$Yb Optical Lattice Clock at NMIJ $IEEE$ $Trans.$ $Ultrason.$ $Ferroelectr.$ $Freq.$ $Control$ $\bf{65}$ 2449-2458
\bibitem{Nemitz2016} Nemitz N, Ohkubo T, Takamoto M, Ushijima I, Das M, Ohmae N and Katori H 2016 Frequency ratio of Yb and Sr clocks with $5\times10^{-17}$ uncertainty at 150 seconds averaging time $Nat.$ $Photon.$ $\bf{10}$ 258-261
\bibitem{Falke2012} Falke S, Misera M, Sterr U, Lisdat C 2012 Delivering pulsed and phase stable light to atoms of an optical clock $Appl.$ $Phys.$ B $\bf{107}$, 301-311
\bibitem{Inaba2013} Inaba H, Hosaka K, Yasuda M, Nakajima Y, Iwakuni K, Akamatsu D, Okubo S, Kohno T, Onae A, and Hong F-L 2013 Spectroscopy of $^{171}$Yb in an optical lattice based on laser linewidth transfer using a narrow linewidth frequency comb $Opt.$ $Express$ $\bf{21}$ 7891-7896 
\bibitem{Nishiyama} Nishiyama A $et$ $al.$ in preparation
\bibitem{Kobayashi2025} Kobayashi T, Nishiyama A, Hosaka K, Akamatsu D, Kawasaki A, Wada M, Inaba H, Tanabe T, and Yasuda M 2025 Improved absolute frequency measurement of $^{171}$Yb at NMIJ with uncertainty below $2\times10^{-16}$ $Metrologia$ $\bf{62}$ 025006
\bibitem{Inaba2006} Inaba H, Daimon Y, Hong F-L, Onae A, Minoshima K, Schibli T R, Matsumoto H, Hirano M, Okuno T, Onishi M, Nakazawa M 2006 Long-term measurement of optical frequencies using a simple, robust and low-noise fiber based frequency comb $Opt.$ $Express$ $\bf{14}$ 5223-5231
 
\bibitem{Nakashima2022} Nakashima M, Fukaya S, Toyofuku T, Ochi K and Matsuo K 2022 Determination of geopotential values at the optical lattice clocks based on geodetic approaches $Japan$ $Geoscience$ $Union$ $Meeting$ p SGD02-14
\bibitem{Beloy2020} Beloy K, McGrew W F, Zhang X, Nicolodi D, Fasano R J, Hassan Y S, Brown R C, and Ludlow A D 2020 Modeling motional energy spectra and lattice light shifts in optical lattice clocks $Phys.$ $Rev.$ A $\bf{101}$ 053416
\bibitem{Zhang2022} Zhang X, Beloy K, Hassan Y S, McGrew W F, Chen C-C, Siegel J L, Grogan T, and Ludlow A D 2022 Subrecoil Clock-Transition Laser Cooling Enabling Shallow Optical Lattice Clocks $Phys.$ $Rev.$ $Lett.$ $\bf{129}$ 113202
\bibitem{Chen2024} Chen C-C, Siegel J L, Hunt B D, Grogan T, Hassan Y S, Beloy K, Gibble K, Brown R C, and Ludlow A D 2024 Clock-Line-Mediated Sisyphus Cooling $Phys.$ $Rev.$ $Lett.$ $\bf{133}$ 053401
\bibitem{Katori2015} Katori H, Ovsiannikov V D, Marmo S I, Palchikov V G 2015 Strategies for reducing the light shift in atomic clocks $Phys.$ $Rev.$ A $\bf{91}$ 052503
\bibitem{Ushijima2018} Ushijima I, Takamoto M, and Katori H 2018 Operational Magic Intensity for Sr Optical Lattice Clocks $Phys.$ $Rev.$ $Lett.$ $\bf{121}$ 263202
\bibitem{Brown2017} Brown R C, Phillips N B, Beloy K, McGrew W F, Schioppo M, Fasano R J, Milani G, Zhang X, Hinkley N, Leopardi H, Yoon T H, Nicolodi D, Fortier T M, and Ludlow A D 2017 Hyperpolarizability and Operational Magic Wavelength in an Optical Lattice Clock $Phys.$ $Rev.$ $Lett.$ $\bf{119}$ 253001
\bibitem{Nemitz2019} Nemitz N, J\o rgensen A A, Yanagimoto R, Bregolin F, and Katori H 2019 Modeling light shifts in optical lattice clocks $Phys.$ $Rev.$ A $\bf{99}$ 033424
\bibitem{Bothwell2025} Bothwell T, Hunt B D, Siegel J L, Hassan Y S, Grogan T, Kobayashi T, Gibble K, Porsev S G, Safronova M S, Brown R C, Beloy K, and Ludlow A D 2025 Lattice Light Shift Evaluations in a Dual-Ensemble Yb Optical Lattice Clock $Phys.$ $Rev.$ $Lett.$ $\bf{134}$ 033201
\bibitem{Pizzocaro2020} Pizzocaro M, Bregolin F, Barbieri P, Rauf B, Levi F, and Calonico D 2020 Absolute frequency measurement of the $^{1}$S$_0$-$^3$P$_0$ transition of $^{171}$Yb with a link to international atomic time $Metrologia$ $\bf{57}$ 035007
\bibitem{Kim2021} Kim H, Heo M-S, Park C Y, Yu D-H and Lee W-K 2021 Absolute frequency measurement of the $^{171}$Yb optical lattice clock at KRISS using TAI for over a year $Metrologia$ $\bf{58}$ 055007
\bibitem{Goti2025} Goti I, Petrucciani T, Condio S, Levi F, Calonico D, Pizzocaro M 2025 Atomic thermometry in optical lattice clocks arXiv:2508.08164
\bibitem{Leibfried2003} Leibfried D, Blatt R, Monroe C, and Wineland D 2003 Quantum dynamics of single trapped ions $Rev.$ $Mod.$ $Phys.$ $\bf{75}$, 281 (2003).
\bibitem{Fasano2021} Fasano R J, Chen Y J, McGrew W F, Brand W J, Fox R W, and Ludlow A D 2021 $Phys.$ $Rev.$ $Appl.$ $\bf{15}$, 044016 (2021). 
\bibitem{Kobayashi2019} Kobayashi T, Akamatsu D, Hosaka K, and Yasuda M 2019 A relocking scheme for optical phase locking using a digital circuit with an electrical delay line $Rev.$ $Sci.$ $Instrum.$ $\bf{90}$ 103002
\bibitem{Beloy2014} Beloy K, Hinkley H, Phillips N B, Sherman J A, Schioppo M, Lehman J, Feldman A, Hanssen L M, Oates C W, and Ludlow A D 2014 Atomic Clock with $1\times10^{-18}$ Room-Temperature Blackbody Stark Uncertainty $Phys.$ $Rev.$ $Lett.$ $\bf{113}$ 260801
\bibitem{Sherman2012} Sherman J A, Lemke N D, Hinkley N, Pizzocaro M, Fox R W, Ludlow A D, and Oates C W 2012 High-Accuracy Measurement of Atomic Polarizability in an Optical Lattice Clock $Phys.$ $Rev.$ $Lett.$ $\bf{108}$ 153002
\bibitem{Hassan2025} Hassan Y S, Beloy K, Siegel J L, Kobayashi T, Swiler E, Grogan T, Brown R C, Rojo T, Bothwell T, Hunt B D, Halaoui A, and Ludlow A D 2025 Cryogenic Optical Lattice Clock with $1.7\times10^{-20}$ Blackbody Radiation Stark Uncertainty $Phys.$ $Rev.$ $Lett.$ $\bf{135}$ 063402
\bibitem{Heo2022} Heo M-S, Kim H, Yu D-H, Lee W-K and Park C Y 2022 Evaluation of the blackbody radiation shift of an Yb optical lattice clock at KRISS $Metrologia$ $\bf{59}$ 055002
\bibitem{Jin2023} Jin T, Zhang T, Luo L, Liu L, Zhou M and Xu X 2023 Multiple strategies for evaluating the uncertainty of blackbody radiation frequency shift in an optical clock $Measurement$ $\bf{216}$ 112946
\bibitem{ITS90} Preston-Thomas H 1990 The International Temperature Scale of 1990 (ITS-90) $Metrologia$ $\bf{27}$ 3
\bibitem{Allison2016} Allison J $et.$ $al.$ 2016 Recent developments in Geant4 $Nucl$. $Instrum$. $Meth$. $Phys$. $Res$. A $\bf{835}$ 186-225 
\bibitem{Ohmae2021} Ohmae N, Takamoto M, Takahashi Y, Kokubun M, Araki K, Hinton A, Ushijima I, Muramatsu T, Furumiya T, Sakai Y, Moriya N, Kamiya N, Fujii K, Muramatsu R, Shiimado T, and Katori H 2021 Transportable Strontium Optical Lattice Clocks Operated Outside Laboratory at the Level of $10^{-18}$ Uncertainty $Adv$. $Quantum$ $Technol$. $\bf{4}$ 2100015
\bibitem{Nicholoson2015} Nicholson T L, Campbell S L, Hutson R B, Marti G E, Bloom B J, McNally R L, Zhang W, Barrett M D, Safronova M S, Strouse G F, Tew W L, and Ye J 2015 Systematic evaluation of an atomic clock at $2\times10^{-18}$ total uncertainty $Nat.$ $Commun.$ $\bf{6}$ 6896
\bibitem{Nicholsonthesis} Nicholson T L 2015 A new record in atomic clock performance Ph.D. thesis, University of Colorado Boulder 
\bibitem{Kobayashi2024} Kobayashi T, Akamatsu D, Hosaka K, Hisai Y, Nishiyama A, Kawasaki A, Wada M, Inaba H, Tanabe T, Suzuyama T, and Yasuda M 2024 Generation of a precise time scale assisted by a near-continuously operating optical lattice clock $Phys.$ $Rev.$ $Appl.$ $\bf{21}$ 064015
\bibitem{recommendedvalue} Recommended values of standard frequencies for applications including the practical realization of the metre and secondary representations of the definition of the second BIPM publication, approved by CCTF September 2025
\bibitem{Wada2022} Wada M and Inaba H 2022 Femtosecond-comb based 10 MHz-to-optical frequency link with uncertainty at the 10$^{-18}$ level $Metrologia$ $\bf{59}$ 065005

\bibitem{Blatt2009} Blatt S, Thomsen J W, Campbell G K, Ludlow A D, Swallows M D, Martin M J, Boyd M M, and Ye J 2009 Rabi spectroscopy and excitation inhomogeneity in a one-dimensional optical lattice clock $Phys.$ $Rev.$ A $\bf{80}$ 052703
\bibitem{Hong2009} Hong F L, Musha M, Takamoto M, Inaba H, Yanagimachi S, Takamizawa A, Watabe K, Ikegami T, Imae M, Fujii Y, Amemiya M, Nakagawa K, Ueda K, and Katori H 2009 Measuring the frequency of a Sr optical lattice clock using a 120 km coherent optical transfer $Opt.$ $Lett.$ $\bf{34}$ 692-694
\bibitem{Akatsuka2020} Akatsuka T, Goh T, Imai H, Oguri K, Ishizawa A, Ushijima I, Ohmae N, Takamoto M, Katori H, Hashimoto T, Gotoh H, and Sogawa T 2020 Optical frequency distribution using laser repeater stations with planar lightwave circuits $Opt.$ $Express$ $\bf{28}$ 9186-9197
\bibitem{Bothwell2025optlett} Bothwell T, Brand W, Fasano R, Akin T, Whalen J, Grogan T, Chen Y-J, Pomponio M, Nakamura T, Rauf B, Baldoni I, Giunta M, Holzwarth R, Nelson C, Hati A, Quinlan F, Fox R, Peil S, and Ludlow A 2025 Deployment of a transportable Yb optical lattice clock $Opt.$ $Lett.$ $\bf{50}$ 646-649
\bibitem{Akamatsu2014} Akamatsu D, Inaba H, Hosaka K, Yasuda M, Onae A, Suzuyama T, Amemiya M, and Hong F-L 2014 Spectroscopy and frequency measurement of the $^{87}$Sr clock transition by laser linewidth transfer using an optical frequency comb $Appl.$ $Phys.$ $Express$ $\bf{7}$ 012401
\bibitem{Hisai2021} Hisai Y, Akamatsu D, Kobayashi T, Hosaka K, Inaba H, Hong F-L, and Yasuda M 2021 Improved frequency ratio measurement with $^{87}$Sr and $^{171}$Yb optical lattice clocks at NMIJ $Metrologia$ $\bf{58}$ 015008


\end{thebibliography}
\end{document}